# Searches for the role of spin and polarization in gravity

Wei-Tou Ni[1,2]

[1]Center for Gravitation and Cosmology, Department of Physics,
National Tsing Hua University, Hsinchu, Taiwan, 30013 ROC
[2] Purple Mountain Observatory, Chinese Academy of Sciences,
Nanjing, 210008 China

E-mail: weitou@gmail.com

**Abstract**

Spin is fundamental in physics. Gravitation is universal. Searches for the role of spin in gravitation dated before the firm establishment of the electron spin in 1925. Since mass and spin or helicity in the case of zero mass are the only invariants of the Poincaré group and mass participates in universal gravitation, these searches are natural steps to pursue. Here we review both the theoretical and experimental efforts in searching for the role of spin/polarization in gravitation. We discuss torsion, Poincaré gauge theories, teleparallel theories, metric-affine connection theories and pseudoscalar (axion) theories. We discuss laboratory searches for electron and nucleus spin-couplings --- the weak equivalence principle experiments for polarized-bodies, the finite-range spin-coupling experiments, the spin-spin coupling experiments and the cosmic-spin coupling experiments. The role played by angular momentum and rotation is explicitly discussed. We discuss astrophysical and cosmological searches for photon polarization coupling. Investigation in the implications and interrelations of equivalence principles led to a possible pseudoscalar or vector interaction, and led to the proposal of WEP II (Weak Equivalence Principle II) which include rotation in the universal free-fall motion. Evidences for WEP II are discussed and compiled. Cosmological searches for photon-polarization coupling test the possibility of violation of EEP and the existence of cosmic pseudoscalar/vector interaction and may reveal a potential influence to our presently-observed universe from a larger arena. In relativistic gravity, there is a Lense-Thirring frame-dragging on rotating body with angular momentum. In analog with gyromagnetic ratio in electromagnetism, one can define gyrogravitational ratio. A profound search for the role of spin in gravitation is to measure the gyrogravitational ratio of particles. This could lead us to probe and understand the microscopic origins of gravity. We discuss the strategies to perform such experiments.

**Contents**

## 1. Introduction

According to our present understanding of physics, particles and fields transform appropriately under inhomogeneous Lorentz transformations. These inhomogeneous Lorentz transformations form a group called the Poincaré group. The only invariants characterizing irreducible representations of the Poincaré group are mass and spin (or helicity in the case of zero mass). Both electroweak and strong interactions are strongly spin-dependent. The question comes whether the gravitational interaction is spin-dependent (polarization-dependent). In this paper, we review the searches for the role of spin in gravitation.

The gravitational interaction is the earliest formulated interaction. Both Newtonian gravitation and Einstein's general relativity are universal interaction theories about masses. There are no polarization-dependent effects in these theories. Historically, these theories were formulated before spin was discovered. Ever since the existence of spin (intrinsic spin) was noticed (before it was firmly established), people started to propose possible polarization-dependent effects in gravitation on various levels. If there are spin-dependent effects in gravitation, Einstein's Equivalence Principle (EEP) would be violated at a certain appropriate level.

Since mass and spin (helicity) are two independent invariants of the Poincaré group, there is the question whether the gravitational interaction between masses, the "gravitational" interaction between masses and spins, and the "gravitational" interaction between spins share the same coupling constant. If the strengths of coupling are different, then the question comes whether we shall call the spin-dependent interaction gravitational. This question can only be answered if the strengths are determined and a working theory is formulated and adopted. From a phenomenological approach, we ask whether there is a long-range (or semi-long-range) spin-mass or spin-spin interaction and what are its strength and its interaction form. Therefore, in reviewing the experimental searches, we include the related efforts.

## 2. Historical background

*2.1. Spin and polarization*

In 1921-22 Stern and Gerlach (Stern 1921; Stern and Gerlach 1922a, b; for a fascinating account of the story of discovery, see Friedrich and Herschbach 2003) discovered the space quantization of atomic magnetic moments. In 1925-26, Uhlenbeck and Goudsmit (1925, 1926) introduced our present concept of electron spin as the culmination of a series of studies of doublet and triplet structures in spectra.

The discovery of the phenomena of light polarization has a long history. In 1669, Bartholinus showed that crystals of 'Iceland spar' (calcite, $CaCO_3$) produced two refracted rays from a single incident beam. Subsequent experiments determined that these two rays possessed unique characteristics as if they have 'sides'. Malus (1809a,b) discovered that reflected light and scattered light also possessed this 'sidesness', which he called 'polarization'. Arago and Fresnel observed that the two polarized beams of light are not to interfere with each other. With the proposal of Young that the oscillations in the optical disturbance were transverse (perpendicular to the direction of propagation), Fresnel formulated his mathematical theory of light which remains useful today (Klein and Furtak 1986). With the advent of quantum theory, we understand that the light polarization also has quantum property and is important in the implementation of quantum measurement and quantum information experiment.

*2.2. Torsion*

In 1921, Eddington (1921) mentioned the notion of an asymmetric affine connection in discussing possible extensions of general relativity. In 1922, Cartan (1922) introduced torsion as the anti-symmetric part of an asymmetric affine connection and laid the foundation of this generalized geometry. Cartan (1923, 1924, 1925) proposed that the torsion of spacetime might be connected with the intrinsic angular momentum of matter.

In local coordinates, the covariant derivative of a contravariant vector field $A^i$ in an affine manifold is defined as

$$A^i_{;k} \equiv A^i_{,k} + \Gamma^i_{jk} A^j, \tag{1}$$

where $i, j, k\ldots$ are coordinate indices, comma denotes partial differentiation and $\Gamma^i_{jk}$ is the affine connection. The Riemann tensor is defined as

$$R^i_{jkl} \equiv \Gamma^i_{jl,k} - \Gamma^i_{jk,l} + \Gamma^i_{mk} \Gamma^m_{jl} - \Gamma^i_{ml} \Gamma^m_{jk}. \tag{2}$$

The Riemann tensor defined in (2) is antisymmetric in the last two indices. [We refer the nonspecialist or graduate student to Misner, Thorne and Wheeler (1973) and Gronwald (1997) for introductory and thorough discussions about the definitions introduced here.]

We can split the connection $\Gamma^i_{jk}$ into its symmetric part $\underline{\Gamma}^i_{jk}$ and its antisymmetric

part $T^i_{jk}$:

$$\Gamma^i_{jk} = \underline{\Gamma}^i_{jk} + T^i_{jk}, \qquad (3)$$

where

$$\underline{\Gamma}^i_{jk} \equiv 1/2\ (\Gamma^i_{jk} + \Gamma^i_{kj}), \qquad (4)$$
$$T^i_{jk} \equiv 1/2\ (\Gamma^i_{jk} - \Gamma^i_{kj}). \qquad (5)$$

$T^i_{jk}$ is called Cartan's torsion tensor or, simply, torsion. Although the affine connection $\Gamma^i_{jk}$ is not a tensor, the torsion $T^i_{jk}$ is a tensor.

*2.3. Poincaré gauge theory, teleparallel theory, metric-affine connection theory*

After the formulation of gauge theory by Yang and Mills in 1954, many efforts have been made to bring the gravitation into the present gauge-theoretic framework.

Utiyama (1956) laid ground work for a gauge theory of gravitation. Combining the ideas of Cartan and gauge formalism, Sciama (1962, 1964) and Kibble (1961) developed a theory of gravitation which is commonly called the Einstein-Cartan-Sciama-Kibble theory (Hehl *et al* 1976). The source for torsion field is spin-density current. However, for this theory, the equation for torsion is algebraic; torsion vanishes in vacuum. To make torsion dynamic, *Poincaré Gauge Theory* (PGT) has been proposed (Hehl 1980, Hayashi and Shirafuji 1980; and references therein) and examined in detail by many authors [see Yo and Nester 1999, 2002; Hammond 2002; de Sabbata and Sivaram 1994; Shapiro 2002; and references therein].

Yang (1974) proposed his gravitational equation in 1974 with a motivation to put gravitation into a gauge theory. However, there are spurious solutions (Ni 1975) and the metric is postulated instead of derived. Affine connections correspond to gauge potentials. To be truly analogous to the present gauge theories, the metric ought to be derivable from the affine connection and the equations of motion. To pursue this approach further, we first obtain the necessary and sufficient conditions for the existence of a metric in an affine manifold (Ni 1981, Cheng and Ni 1980). Now the problem comes as how to transform these conditions into equations of motion derivable from a variational principle. Ashtekar's (1986, 1987, 1991) formulation of general relativity is an approach in this general direction.

Two basic subjects for gravity are the tetrad $e_a^i$ and the affine connection $\Gamma^i_{jk}$; tetrad determines the (symmetric) metric $g^{ij}$ (= $\eta^{ab} e_a^i e_b^j$) and the locally Lorentz frame, while affine connection defines the parallel transport and covariant derivative. Here the tetrad indices $a$, $b$, $c$… run from 0 to 3 and are raised and lowered by Minkowski metric $\eta^{ab} \equiv \text{dia}(1, -1, -1, -1)$. Tetrad (or metric) and affine connection are two independent mathematical objects. In gravitation, we seek to find their relation. The covariant derivative of the metric is called the nonmetricity $Q_k{}^{ij}$:

$$Q_k{}^{ij} \equiv g^{ij}{}_{;k}. \tag{6}$$

With the definition nonmetricity and torsion, one can show that

$$\Gamma^i{}_{jk} = \{^i{}_{jk}\} + T^i{}_{jk} + T_k{}^i{}_j + T_j{}^i{}_k + \tfrac{1}{2}(Q^i{}_{jk} + Q^i{}_{kj} - Q_k{}^i{}_j) \tag{7}$$

where the Christoffel symbol $\{^i{}_{jk}\}$ is defined by

$$\{^i{}_{jk}\} \equiv \tfrac{1}{2} g^{im}(g_{jm,k} + g_{mk,j} - g_{jk,m}). \tag{8}$$

Manifold with a metric and the affine connection given by the Christoffel symbol constructed from the metric is called a Riemann manifold. If the affine connection is given independently but satisfies the compatibility condition

$$Q_k{}^{ij} \equiv g^{ij}{}_{;k} = 0, \tag{9}$$

this manifold is called a Riemann-Cartan manifold. In a Riemann or Riemann-Cartan manifold, the metric is used to raise or lower the indices. In a Riemann-Cartan manifold, the only independent degrees of freedom of the affine connection are the torsion degrees of freedom and the affine connection is related to Christoffel symbol by the following equation

$$\Gamma^i{}_{jk} = \{^i{}_{jk}\} + K^i{}_{jk}, \tag{10}$$

with the contortion $K^i{}_{jk}$ defined by

$$K^i{}_{jk} = T^i{}_{jk} + T_k{}^i{}_j + T_j{}^i{}_k. \tag{11}$$

The torsion tensor $T^i{}_{jk}$ (defined in Eq. (5)) can be decomposed into its Lorentz irreducible parts: a vector $v_i$, an axial vector (pseudovector) $a_i$, and an irreducible tensor which is traceless and symmetric with respect to the first 2 indices $t_{ijk}$ defined as follows

$$v_i = T^j{}_i{}^j, \tag{12}$$

$$a_i = (1/6)\,\varepsilon_{ijkl}\,T^{jlk}, \tag{13}$$

$$t_{ijk} = -T_{(ij)k} + (1/6)(g_{ik}\,v_j + g_{jk}\,v_i) - (1/3)\,g_{ij}\,v_k, \tag{14}$$

where the completely antisymmetric tensor $\varepsilon_{ijkl}$ in (13) is defined as

$$\varepsilon_{ijkl} = g_{mi}\,g_{nj}\,g_{pk}\,g_{ql}\,\varepsilon^{mnpq}; \quad \varepsilon^{ijkl} = (-g)^{-1/2}\,e^{ijkl}, \tag{15}$$

with $g$ the determinant of $g_{ij}$ and the antisymmetric symbol $e^{ijkl}$ defined as

$$e^{ijkl} = \begin{cases} 1, & \text{if } (ijkl) \text{ is an even permutation of } (0123) \\ -1, & \text{if } (ijkl) \text{ is an odd permutation of } (0123), \\ 0, & \text{otherwise.} \end{cases} \tag{16}$$

The completely antisymmetric tensor $\varepsilon_{ijkl}$ is a pseudotensor under local P (parity) and T (time reversal) transformation. Hence, $a_i$ defined in (13) is a pseudovector.

In the Poincaré gauge theory, these irreducible parts are used to construct the Lagrangian which is quadratic in the Ricci curvature and these irreducible parts. In general, it has ten parameters. They are analyzed in detail in many research works (See Yo and Nester, 1999, 2002 and references therein). For some of the parameters, the source of torsion can be ordinary angular momentum, not just intrinsic spin angular momentum. In these cases, the torsion is experimentally measurable (See also Dereli and Tucker, 1982, 2002 and references therein). In fact, there are various torsion cosmological models trying to take account of the supernova acceleration observation (Huang *et al* 2008, Shie *et al* 2008, Li *et al* 2009; and references therein). Inflationary models with torsion have also been attempted (Wang and Wu 2009; and references therein).

If we use the minimally coupled Dirac Lagrangian for spin 1/2 particle (Hehl, von der Heyde *et al* 1976; Lämmerzahl 1997), the Dirac equation is

$$i\hbar\gamma^i D_i\psi + (i/2)(K_{ji}{}^j\gamma^i\psi) + mc\psi = i\hbar\gamma^i \underline{D}_i\psi - \hbar a_i\gamma_5\gamma^i\psi + mc\psi = 0, \tag{17}$$

where

$$D_j\psi = \partial_i\psi + \Gamma_i\psi, \tag{18}$$

with the spinorial representation of the anholonomic connection

$$\Gamma_i = (1/4)\, D_i e_a{}^k e^b{}_k \gamma_b \gamma^a = (1/4)\, \underline{D}_i e_a{}^k e^b{}_k \gamma_b \gamma^a - (1/4)\, K^l{}_{ik} e_a{}^k e^b{}_l \gamma_b \gamma_5 \gamma^a. \qquad (19)$$

$\underline{D}_i$ is the Christoffel part of the covariant derivative and the axial vector $a_i$ as defined by (13) is the axial-vector part of the space-time torsion. By analyzing Hughes-Drever experiments in this context, Lämmerzahl (1997) obtained constraint on the axial torsion $|a_\alpha| \leq 1.5 \times 10^{-15}$ m$^{-1}$ where $|a_\alpha|\ [= (a_1{}^2 + a_2{}^2 + a_3{}^2)^{1/2}]$ is the absolute value of the spatial part of the axial torsion.

Teleparallel theory assumes there is a global parallel tetrad. A Riemann-Cartan manifold with a global parallel tetrad is called Weitzenböck space. In this space, there is no curvature, i.e., Riemann curvature is zero and there is an absolute parallelism. The New General Relativity of Hayashi and Shirafujii (1979) is such a theory. In this theory, torsion is generated by both spin and angular momentum.

Metric-affine theories (MAGs) treat metric and affine connection more or less on equal footing. MAG theories use $g$, $T$ and $Q$ to construct Lagrangians. As an example, the Lagrangian of Hehl, Kerlick and von der Heyde (1976) theory is written as

$$L = (-g)^{1/2}\, g^{ij}[R_{ij}(\Gamma, \partial \Gamma) + \beta Q_i Q_j], \qquad (20)$$

where the Ricci tensor $R_{ij}$ is considered as a function of affine connection and its derivative, $\beta$ is a dimensionless coupling constant, and $Q_i$ $(= -(1/4)\, Q^j{}_{ij})$ is proportional to a trace of the nonmetricity $Q_{ijk}$. There are various studies of MAG theories (e.g., Tucker and Wang 1995, Dereli *et al* 1996) many of them are reviewed in Hehl *et al* (1995), Gronwald (1997), Hehl and Macias (1999; and references therein).

*2.4. Pseudoscalar term and pseudoscalar theories*

In 1973, we studied the relationship of Galilio Equivalence Principe (WEP I) and Einstein Equivalence Principle in a framework (the χ-g framework [see section 3.1]) of electromagnetism and charged particles, we found the following example with interaction Lagrangian density

$$\boldsymbol{L}_{\text{int}} = -(\frac{1}{16\pi})(-g)^{1/2}[\frac{1}{2}g^{ik}g^{jl} - \frac{1}{2}g^{il}g^{kj} + \varphi\varepsilon^{ijkl}]F_{ij}F_{kl} - A_k j^k (-g)^{1/2} - \sum_I \frac{ds_I}{dt}\delta(x - x_I)$$
$$, \qquad (21)$$

as an example which obeys WEP I, but not EEP (Ni 1973, 1974, 1977). Here $\varphi$ is a scalar or pseudoscalar function of relevant variables. If we assume that the $\varphi$-term is local CPT invariant, then $\varphi$ should be a pseudoscalar (function) since $\varepsilon^{ijkl}$ is a pseudotensor. The nonmetric part of this theory is

$$L^{(NM)}{}_{int} = -(\frac{1}{16\pi})(-g)^{1/2}\varphi\varepsilon^{ijkl}F_{ij}F_{kl} = -(\frac{1}{4\pi})(-g)^{1/2}\varphi_{,i}\,\varepsilon^{ijkl}A_jA_{k,l} \quad \text{(mod div)}, \qquad (22)$$

where 'mod div' means that the two Lagrangian densities are related by integration by parts in the action integral. The Maxwell equations (Ni 1973, 1977) become

$$F^{ik}{}_{|k} + \varepsilon^{ikml}F_{km}\varphi_{,l} = -4\pi j^i, \qquad (23)$$

where the derivation | is with respect to the Christoffel connection. The Lorentz force law is the same as in metric theories of gravity or general relativity. Gauge invariance and charge conservation are guaranteed. The modified Maxwell equations (23) are also conformally invariant.

This theory can be put into the form of a torsion theory. Define a metric compatible affine connection as in (10) with the contorsion defined by

$$K^i{}_{jk} = 2\varphi_{,l}\varepsilon^{li}{}_{jk}. \qquad (24)$$

We note that with this definition the contorsion $K^i{}_{jk}$ is equal to torsion $T^i{}_{jk}$. The Modified Maxwell equations (23) can then be written as

$$\underline{F}^{ik}{}_{;k} = -4\pi j_i \qquad (25)$$

where ";" denotes covariant differentiation with respect to the affine connection $\Gamma^i{}_{jk}$ and

$$\underline{F}_{ik} \equiv A_{k;i} - A_{i;k} = A_{k,i} - A_{i,k} + 2T^l{}_{ik}A_l. \qquad (26)$$

The nonmetric part of the Lagrangian can be written in the form (Ni 1983c)

$$L^{(NM)}{}_I = 2A_jA_{k,l}T^{jkl}(-g)^{1/2}. \qquad (27)$$

To complete this theory as a gravitational theory, we have to add a gravitational Lagrangian to it. For example, the gravitational Lagrangian $L_G$ could be

$$L_G = (1/16\pi)\times(-g)^{1/2}R(\Gamma^i{}_{jk}), \qquad (28)$$

$$L_G = (1/16\pi)\times(-g)^{1/2} [R(\Gamma^i_{jk}) + \eta\varphi_{,i}\varphi^{,i}], \tag{29}$$

or

$$L_G = (1/16\pi)\times(-g)^{1/2} [\varphi R(\{^i_{jk}\}) - (1/\varphi)\omega(\varphi)\, \varphi_{,i}\varphi^{,i}], \tag{30}$$

where η is a parameter and ω(φ) is a function of φ (Ni 1983c). Various different extensions have been considered, many of them are reviewed in Balakin and Ni (2009).

If we add the Dirac Lagrangian density, we obtain equation (17) as the Dirac equation of this theory with $a_i$ equals to $-2\varphi_{,i}$. As we saw in the last subsection, this part of interaction is constrained.

The rightest term in equation (22) is reminiscent of Chern-Simons (1974) term $e^{\alpha\beta\gamma} A_\alpha F_{\beta\gamma}$. There are two differences: (i) Chern-Simons term is in 3 dimensional space; (ii) Chern-Simons term in the integral is a total divergence. However, it is interesting to notice that the cosmological time may be defined through the Chern-Simons invariant (Smolin and Soo 1995).

A term similar to the one in equation (22) (axion-gluon interaction) occurs in QCD in an effort to solve the strong CP problem (Peccei and Quinn 1977, Weinberg 1978, Wilczek 1978). Carroll, Field and Jackiw (1990) proposed a modification of electrodynamics with an additional $e^{ijkl} V_i A_j F_{kl}$ term with $V_i$ a constant vector (See also Jackiw 2007). This term is a special case of the term $e^{ijkl} \varphi\, F_{ij} F_{kl}$ (mod div) with $\varphi_{,I} = -\tfrac{1}{2}V_i$.

Various terms in the Lagrangians discussed in this subsection are listed in Table 1. Empirical tests of the pseudoscalar-photon interaction (22) will be discussed in section 5 together with related theoretical models.

**Table 1.** Various terms in the Lagrangian and their meaning.

| Term | Dimension | Reference | Meaning |
| --- | --- | --- | --- |
| $e^{\alpha\beta\gamma} A_\alpha F_{\beta\gamma}$ | 3 | Chern-Simons (1974) | Intergrand for topological invarinat |
| $e^{ijkl} \varphi\, F_{ij} F_{kl}$ | 4 | Ni (1973, 1974, 1977) | Pseudoscalar-photon coupling |
| $e^{ijkl} \varphi\, F^{QCD}_{ij} F^{QCD}_{kl}$ | 4 | Peccei-Quinn (1977) Weinberg (1978) Wilczek (1978) | Pseudoscalar-gluon coupling |
| $e^{ijkl} V_i A_j F_{kl}$ | 4 | Carroll-Field-Jackiw (1990) | External constant vector coupling |

## 3. Theoretical connections and motivations

To look for theoretical connections, we first examine the long standing equivalence principles in theoretical frameworks, both theoretically and empirically, and give motivations to test them further with comments on the origin of equivalence. Related to equivalence, we discuss inertial forces and macroscopic manifestation of inertial torques on intrinsic spins. We then discuss proposed interactions with spin or polarization dependence. Finally in this section we discuss the relevance of gyro-gravitational effects to the microscopic origin of gravity and some promising methods to measure the gyro-gravitational ratios of elementary particles.

*3.1. WEP I, WEP II, EEP and the pseudoscalar-photon interaction*

Equivalence principles (Galilei 1683, Einstein 1907) are cornerstones in the foundation of gravitation theories. In the theoretical study of the foundation problems, to what extent the Galileo weak equivalence principle [Universality of free-fall trajectories (WEP I) implies the validity of the Einstein equivalence principle (EEP) is an important issue. *EEP, as precisely stated by Misner, Thorne and Wheeler (1973), is that in the locality of every point (event) in spacetime, the nongravitational physics is that of special relativity.* Schiff (1960) conjectured that the Galileo weak equivalence principle implies the Einstein equivalence principle. In 1972, we started to investigate this issue and reached a counterexample of Schiff's conjecture (Ni 1973). In order to find out to what extent the violation occurs, we followed up using a general framework --- the $\chi$-g framework to study Schiff's conjecture and theoretical relations of various equivalence principles. This counterexample remains the only example in the $\chi$-g framework that violates the Schiff's conjecture (Ni 1974, 1977).

The $\chi$-g framework can be summarized in the following interaction Lagrangian density

$$L_{\text{int}} = -(\frac{1}{16\pi})\chi^{ijkl} F_{ij} F_{kl} - A_k j^k (-g)^{1/2} - \sum_I \frac{ds_I}{dt} \delta(x - x_I), \tag{31}$$

where $\chi^{ijkl} = \chi^{klij} = -\chi^{klji}$ is a tensor density of the gravitational fields (e.g., $g_{ij}$, $\phi$, etc.) or fields to be investigated and $j^k$, $F_{ij} \equiv A_{j,i} - A_{i,j}$ have the usual meaning in electromagnetism. The gravitation constitutive tensor density $\chi^{ijkl}$ dictates the behavior of electromagnetism in a gravitational field and has 21 independent components in general. For a metric theory (when EEP holds), $\chi^{ijkl}$ is determined completely by the metric $g^{ij}$ and equals $(-g)^{1/2}(\frac{1}{2} g^{ik} g^{jl} - \frac{1}{2} g^{il} g^{kj})$.

We proved that for a system whose Lagrangian density is given by Eq. (1), WEP I holds if and only if

$$\chi^{ijkl} = (-g)^{1/2}[\frac{1}{2}g^{ik}g^{jl} - \frac{1}{2}g^{il}g^{kj} + \varphi\varepsilon^{ijkl}], \qquad (32)$$

where $\phi$ is a scalar function of the gravitational fields or fields to be investigated, and $\varepsilon^{ijk}$ is defined in equations (15) and (16). We have discussed this theory in subsection 2.4. in the context of historical background. Now we discuss it in relation to equivalence principles.

If $\phi \neq 0$ in (2), the gravitational coupling to electromagnetism is not minimal and EEP is violated. Hence WEP I does not imply EEP and Schiff's conjecture is incorrect (Ni 1973, 1974, 1977). However, WEP I does constrain the 21 degrees of freedom of χ to only one degree of freedom (φ), and Schiff's conjecture is largely right in spirit.

The theory with $\varphi \neq 0$ is a pseudoscalar theory with important astrophysical and cosmological consequences (section 5). This is an example that investigations in fundamental physical laws lead to implications in cosmology. Investigations of CP problems in high energy physics leads to a theory with a similar piece of Lagrangian with φ the axion field for QCD (Peccei and Quinn 1977, Weinberg 1978, Wilczek 1978, Kim 1979, Shifman *et al* 1980, Dine *et al* 1981, Cheng *et al* 1995).

In the nonmetric theory with $\chi^{ijkl}$ ($\varphi \neq 0$) given by Eq. (2) (Ni 1973, 1974, 1977), there are anomalous torques on electromagnetic-energy-polarized bodies so that different test bodies will change their rotation state differently, like magnets in magnetic fields. Since the motion of a macroscopic test body is determined not only by its trajectory but also by its rotation state, the motion of polarized test bodies will not be the same. We, therefore, have proposed the following stronger weak equivalence principle (WEP II) to be tested by experiments, which states that in a gravitational field, both the translational and rotational motion of a test body with a given initial motion state is independent of its internal structure and composition (universality of free-fall motion) (Ni 1974, Ni 1977). To put in another way, the behavior of motion including rotation is that in a local inertial frame for test-bodies. If WEP II is violated, then EEP is violated. Therefore from above, in the χ-g framework, the imposition of WEP II guarantees that EEP is valid.

WEP II state that the motion of all six degrees of freedom (3 translational and 3 rotational) must be the same for all test bodies as in a local inertial frame. There are two different scenarios that WEP II would be violated: (i) the translational motion is affected by the rotational state; (ii) the rotational state changes with angular momentum (rotational direction/speed) or species.

In the latter part of 1980's and early 1990's, a focus is on whether the rotation state would affect the trajectory. Soon after Hayasaka and Takeuki (1989) reported their results that in weighing gyros, gyros with spin vector pointing downward reduced weight proportional to their rotational speed while gyros with spin vector pointing upward did not change weight. This would be a violation of WEP II if confirmed. Since the change in weight $\delta m$ is proportional to the angular momentum in this experiment, the violation could be characterized by the parameter ν defined to be $\delta m/I$ where $I$ is the angular momentum of the gyro. All the experiments by other groups followed did not confirm the report of Hayasaka and Takeuchi (1989).

Table 2 compiles the experimental results. In the second and third column, we list the parameter ν and the Eötvös parameter $\eta$ measured in each experiment. The Eötvös parameter $\eta$ is defined as $\delta m/m$. The angular momentum $I$ is given by $I = 2\pi f\, m\, r_{gyration}^2$ where $r_{gyration}$ (= [moment of inertia/$m$]$^{1/2}$) is the radius of gyration for the rotating body. Hence, $\nu = \eta / (2\pi f\, r_{gyration}^2)$. For rotating bodies, GP-B experiment (Everitt *et al* 2008) has the best accuracy; it is 3-4 orders better than the second best experimental result for rotating bodies. In calculating the ν and $\eta$ parameters for GP-B, we use the data listed in the Gravity Probe B Quick Facts (Gravity Probe B --- Post Flight Analysis Final Report 2007): gyroscope size, 3.81 cm; spin rate, between 5000-10000 rpm. There are four gyroscopes with one of them also as a drag-free test body. The drag-free performance is better than $10^{-11}$ g. In a more precise analysis, the relative acceleration of different gyros with different speed needs to be deduced from levitating feedback data and local space gravity distribution. Here we simple take $10^{-11}$ g as an upper bound of the Eötvös parameter $\eta$. With its precision, GP-B gives a constraint on ν much better than others.

Table 2. Test of WEP II regarding to trajectory using bodies with different angular momentum. The last two rows are for electron spins.

| Experiment | ν [s/cm$^2$] | \|η\| | Method |
| --- | --- | --- | --- |
| Hayasaka-Takeuchi (1989) | $(-9.8\pm0.9)\times10^{-9}$ for spin up, $\pm0.5\times10^{-9}$ for spin down | up to $6.8\times10^{-5}$ | weighing |
| Faller *et al* (1990) | $\pm4.9\times10^{-10}$ | $< 9\times10^{-7}$ | weighing |
| Quinn-Picard (1990) | $\|\nu\| \leq 1.3\times10^{-10}$ | $< 2\times10^{-7}$ | weighing |
| Nitschke-Wilmarth (1990) | $\|\nu\| \leq 1.3\times10^{-10}$ | $< 5\times10^{-7}$ | weighing |
| Imanishi *et al* (1991) | $\|\nu\| \leq 5.8\times10^{-10}$ | $< 2.5\times10^{-6}$ | weighing |
| Luo *et al* (2002) | $\|\nu\| \leq 3.3\times10^{-10}$ | $\leq 2\times10^{-6}$ | free-fall |
| Zhou *et al* (2002) | $\|\nu\| \leq 2.7\times10^{-11}$ | $\leq 1.6\times10^{-7}$ | free-fall |
| Everitt *et al* (2008) | $6.6\times10^{-15}$ | $\leq 1\times10^{-11}$ | free-fall |
| Ni *et al* (1990) | $\|v_{spin}\| \leq 8.6\times10^{-3}$<br>$\|v_{orbit}\| \leq 4.3\times10^{-3}$<br>$\|v_{total}\| \leq 8.6\times10^{-3}$ | $\leq 5\times10^{-9}$ | weighing |
| Hou-Ni (2001) | $\|v_{spin}\| \leq 14.7\times10^{-3}$<br>$\|v_{orbit}\| \leq 8.3\times10^{-3}$<br>$\|v_{total}\| \leq 14{,}7\times10^{-3}$ | $\leq 7.1\times10^{-9}$ | torsion balance |

Results for quantum spin angular momentum from weighing polarized-bodies (Ni 1990) and from polarized-body torsion balance experiment are also listed. Static macroscopic polarized-body has a net spin/orbital angular momentum/total angular momentum. Therefore as far as angular momentum is concerned, it is an invisible rotor. For our shielded polarized bodies, the magnetic moment is compensated. Since the electron spin gyromagnetic factor is twice its orbital gyromagnetic factor and nuclear polarization is small, the net total angular momentum is twice the spin angular momentum and in the opposite direction (Ni 1986, Hou *et al* 2000). The total angular momentum is equal to the spin angular momentum but in opposite direction.

To test WEP II regarding to the rotational state changes with different angular momentum (rotational direction/speed) or species, one needs to measure the rotational direction and speed very precisely with respect to time. GP-B has four gyros rotating with different speeds and has measured the rotational directions very precisely. The quartz rotors have been placed in high-vacuum housing with a very long spin-down rate. We define a WEP II violation parameter $\lambda$ for a test body to be the anomalous torque on the rotating body divided by its angular momentum. Anomalous torque is equal to anomalous angular momentum change divided by time. Angular momentum change divided by angular momentum and time is angular drift in the transverse (to rotation axis) direction and rate of change of the rotation speed in the axial direction. For GP-B, the rotation is in the direction of the guide star IM Pegasi (HR 8703). GP-B experiment is discussed in 4.2.2. We list the constraints on $\lambda$ in all three directions in Table 3. The GP-B results agree with General Relativity, we take their current $2\sigma$ as our preliminary estimate.

Table 3. Test of WEP II regarding to rotational state using rotating quartz balls from GP-B experiment.

| Constraints on the WEP II violation parameter $\lambda$ from GP-B experiment | |
|---|---|
| constraint in the direction of geodetic effect | $\|\lambda\| < 4.3\times10^{-15}$ s$^{-1}$ |
| constraint in the direction of frame dragging effect | $\|\lambda\| < 1.8\times10^{-15}$ s$^{-1}$ |
| constraint in the direction of guide star | $\|\lambda\| < 3\times10^{-11}$ s$^{-1}$ |

In the next subsection, we will discuss the macroscopic manifestation of inertial torques on the spin. This is related to the tests of WEP II.

*3.2. Macroscopic manifestation of inertial torques on the spin*

In the earth-bound laboratories, gyros or bodies with angular momentum experience inertial torques due to earth rotation (Ni and Zimmermann 1978, and references therein). Spin polarized bodies have been made for performing gravity experiments to probe the role of spin in gravitation. If the intrinsic quantum spins and the ordinary angular momenta are mechanically equivalent, these spin-polarized bodies would experience inertial torques too. The analysis of inertial torques on spin-polarized bodies in various experiment have been presented in 1984 (Ni 1984c). Detection of the inertial torques would give macroscopic manifestation of quantum spins. According to EEP, there should also be a correspondence in gravity.

The effective Hamiltonian for a body with ordinary angular momentum **L** in an earth-bound laboratory is $H_{eff} = -\mathbf{\Omega} \cdot \mathbf{L}$ where **Ω** is the angular velocity of the earth rotation. This inertial effect can be called angular momentum-rotation 'coupling'. Assuming intrinsic spin angular momentum is equivalent to ordinary angular momentum in mechanics, the effective Hamiltonian for intrinsic spin is $H_{eff} = -\mathbf{\Omega} \cdot \mathbf{S}$ on earth. This is the spin-rotation coupling purported by Mashhoon (1988). The effective Hamiltonian due to total angular momentum **J** (= **L** + **S**) is $H_{eff} = -\mathbf{\Omega} \cdot \mathbf{J}$. As discussed in the last subsection, for our polarized bodies, **J** = − **S** = ½**L**. The cosmic-spin coupling experiments search for a spin interaction of the type $H_{eff} = \mathbf{C} \cdot \mathbf{\sigma} = C_1\sigma_1 + C_2\sigma_2 + C_3\sigma_3 = 2 \mathbf{C} \cdot \mathbf{S}$ for electrons. Hence this kind of experiments would be able to detect the inertial spin effect (Hou *et al* 2000). We could define a parameter ζ to denote the ratio of anomalous (deviation) effect to the inertial spin effect. The inertial effect including that of the orbital angular momentum is calculated to be equivalent to a $C_3$ of $2.4 \times 10^{-20}$ eV. For our cosmic-spin coupling experiment (Hou *et al* 2003), the bound on $C_3$ (earth rotation direction) is $7 \times 10^{-19}$ eV. The bound on |ζ| is therefore 30.

Heckel *et al* (2006, 2008) performed an improved cosmic-spin coupling experiment and measured the inertial spin effect with orbit angular momentum (they called it gyro-compass effect). They use this measurement to determine the spin content of the polarized body (5 % precision) in agreement with other independent measurement to 18 %. Putting in another way, the inertial spin effect is experimentally confirmed to 0.18.

Experiment of Heckel *et al* (2006, 2008) demonstrated a number of important things: (i) mechanical manifestation of microscopic angular momentum; (ii) intrinsic spin angular momentum is equivalent to orbital angular momentum mechanically; (iii) when used to measure net spin and magnetic field, it can be more precise than ordinary methods in condensed matter measurement.

Spin inertial effect in the lab frame is subtracted in the search for anomalous spin-dependent forces using stored-ion spectroscopy of Wineland e*t al* (1991). From their uncertainty quoted it is about 1 in ζ. In their nuclear spin gyroscope based on an atomic comagnetometer, Kornack *et al* (2005) measured the earth rotation to 3 % using the spin inertial effect, that is |ζ| ≤ 0.03.

We compile these results in Table 4.

Table 4. Manifestation of inertial effects on the intrinsic spins and test of WEP II.

| Experiment | |ζ| | Method |
|---|---|---|
| Wineland e*t al* (1991) | ≤ 1 | NMR |
| Hou e*t al* (2000, 2003) | ≤ 30 | Torsion balance with a polarized-body |
| Kornack e*t al* (2005) | ≤ 0.03 | Nuclear comagnetometer |
| Heckel (2006, 2008) | ≤ 0.18 | Torsion balance with a polarized-body |

*3.3. Origin of equivalence (Ni 1991)*

So far, every experiment confirms the equivalence principle. What is the microscopic origin of the equivalence?

The standard answer to this question is that there is only one tensor field (metric) which mediates gravitation. This gives minimal coupling and equivalence.

To have a working criterion for testing equivalence principles and to explore serious possibilities, I sometimes have a different point of view on the origin of the equivalence:

THE ORIGIN OF EQUIVALENCE IS IDENTITY!

Let us explain this point of view by looking at the significance of the precision Eötvös-Dicke-Braginsky type experiments. The most recent experiments verify the Universality of Free Fall (Galileo Equivalence Principle) for ordinary matter to $10^{-12}$-$10^{-13}$ precision (Schlamminger *et al* 2008). If we want to know how quarks and gluons are equivalent in a gravitational field, we need to know the quark and gluon contents of nucleons and nuclei. An estimate (Weinberg 1977) of quark masses is $m_u$ = 4.2 MeV, $m_d$ = 7.5 MeV, $m_s$ = 150 Mev with $m_d$ - $m_u$ = 3.3 MeV. The neutron-proton mass difference is 1.3 MeV. From these, the differences in electric-type gluon energy, magnetic-type gluon energy and quark energy would be of the order of 0.1%. From high-energy experiments with nuclear targets, one concludes that these differences can be enhanced for nuclei. Differences up to 1% are reasonable. In contrast to the usual thinking that the strong interaction obeys the Universality of Free Fall to as good as experimental accuracy, we actually lose a factor of 100 to 1000, because ordinary matter has quite similar gluon and quark energy contents (Ni 1987). So, empirically, we only know the equivalence of up quark, down quark and gluons to $10^{-10}$.

If quarks were made from "pre-quark", the pre-quark contents might be rather similar and the equivalence of pre-quarks might be lower, say $10^{-6}$. If we went on this way, the pre-pre-pre-quarks might not be equivalent at all. Therefore to the extent things are identical, they are equivalent and the origin of equivalence is identity.

This criterion is useful in searching for experiments to test the equivalence principle. For example, the strange quark mass is very different from the up and down quark masses and the generation of mass is beyond the QCD. Therefore strange matter and Higgs are much more different from the ordinary matter. A test of it for equivalence would be significant.

*3.4. Theoretical frameworks*

In this section, we discuss various theoretical frameworks. We start by looking at experimental constraints in the $\chi$-$g$ framework. This is an effort to experimentally search for electromagnetic polarization coupling and tests of EEP. The most tested part of equivalence is the Galilio equivalence principle (WEP I) on unpolarized bodies. We have also reviewed the present tests on WEP II in subsection 3.1. Since the polarized electromagnetic energy contents of laboratory polarized-bodies and unpolarized-bodies are small, other experimental and observational evidences are crucial in laying the foundation for the Einstein equivalence principle. In the following, we discuss the constraints from all these evidences and how to generalize the $\chi$-$g$ framework to give a more general framework for testing the foundation of relativistic gravity including microscopic phenomena.

In the $\chi$-$g$ framework, for a weak gravitational field,

$$\chi^{ijkl} = \chi^{(0)ijkl} + \chi^{(1)ijkl} \tag{33}$$

where

$$\chi^{(0)ijkl} = (1/2)\eta^{ik}\eta^{jl} - (1/2)\eta^{il}\eta^{kj} \tag{34}$$

with $\eta^{ij}$ the Minkowski metric and $|\chi^{(1)}\text{'s}| \ll 1$. In this field the dispersion relation for $\omega$ for a plane-wave propagating in the z-direction is

$$\omega_\pm = k\{1+(1/4)[(K_1+K_2) \pm \sqrt{(K_1-K_2)^2 + 4K^2}\,]\} \tag{35}$$

where

$$\begin{aligned}K_1 &= \chi^{(1)1010} - 2\chi^{(1)1013} + \chi^{(1)1313}, \\ K_2 &= \chi^{(1)2020} - 2\chi^{(1)2023} + \chi^{(1)2323}, \\ K &= \chi^{(1)1020} - \chi^{(1)1023} - \chi^{(1)1320} + \chi^{(1)1323}.\end{aligned} \tag{36}$$

Photons with two different polarizations propagate with different speeds $V_\pm = \omega_\pm/k$ and would split in 4-dimensional spacetime. The conditions for no splitting (no retardation) is $\omega_+ = \omega_-$, i.e.,

$$K_1 = K_2, \qquad K = 0. \tag{37}$$

Eq. (37) gives two constraints on the $\chi^{(1)}\text{'s}$ (Ni 1983a, 1984a,b).

*Constraints from no birefringence.* The condition for no birefringence (no splitting, no retardation) for electromagnetic wave propagation in all directions in the weak field limit gives ten constraint equations on the $\chi$'s. With these ten constraints, $\chi$ can be written in the following form

$$\chi^{ijkl} = (-H)^{1/2}[(1/2)H^{ik}H^{jl} - (1/2)H^{il}H^{kj}]\psi + \varphi e^{ijkl}, \tag{38}$$

where $H = \det(H_{ij})$ and $H_{ij}$ is a metric which generates the light cone for electromagnetic propagation (Ni 1983a, 1984a,b). Note that (38) has the same form as (32) with $g^{ij}$ replaced by $H^{ij}$ and with an added factor, 'dilation', $\psi$. Recently, Lämmerzahl and Hehl (2004) have shown that this non-birefringence guarantees, without approximation, Riemannian light cone, i.e., Eq. (38) holds without the assumption of weak field. To fully recover EEP, we need (i) good constraints from no birefringence, (ii) good constraints on no extra physical metric, (iii) good constraints on no $\psi$ ('dilaton'), and (iv) good constraints on no $\varphi$ (axion) or no pseudoscalar-photon interaction.

Eq. (38) is verified empirically to high accuracy from pulsar observations and from polarization measurements of extragalactic radio sources. With the null-birefringence observations of pulsar pulses and micropulses before 1980, the relations (38) for testing EEP are empirically verified to $10^{-14} - 10^{-16}$ (Ni 1983a, 1984a, 1984b). With the present pulsar observations, these limits would be improved; a detailed such analysis is given by Huang (2002). Analyzing the data from polarization measurements of extragalactic radio sources, Haugan and Kauffmann (1995) inferred that the resolution for null-birefringence is 0.02 cycle at 5 GHz. This corresponds to a time resolution of $4 \times 10^{-12}$ s and gives much better constraints. With a detailed analysis and more extragalactic radio observations, (38) would be tested down to $10^{-28}$-$10^{-29}$ at cosmological distances. In 2002, Kostelecky and Mews (2002) used polarization measurements of light from cosmologically distant astrophysical sources to yield stringent constraints down to $2 \times 10^{-32}$. The electromagnetic propagation in Moffat's nonsymmetric gravitational theory (Moffat 1991, Cornish *et al* 1995) fits the $\chi$-$g$ framework. Krisher (1991), and Haugan and Kauffmann (1995) have used the pulsar data and extragalactic radio observations to constrain it. It is interesting to note that just as the $\chi$-$g$ framework (Ni, 1974, 1977) was being developed, an upper limit was set for polarization effects on gravitational deflection for radio waves passing through Sun's gravitational field (Harwit *et al* 1974). In the remaining part of this subsection, we assume (38).

*Constraints on One Physical Metric and no 'Dilation' ($\psi$).* Let us now look into the empirical constraints for $H^{ij}$ and $\psi$. In Eq. (31), $ds$ is the line element determined from the metric $g_{ij}$. From Eq. (38), the gravitational coupling to electromagnetism is determined by the metric $H_{ij}$ and two (pseudo)scalar fields $\varphi$ 'axion' and $\psi$ 'dilaton'. If $H_{ij}$ is not proportional to $g_{ij}$, then the hyperfine levels of the lithium atom, the beryllium atom, the mercury atom and other atoms will have additional shifts. But this is not observed to high accuracy in Hughes-Drever experiments (Hughes *et al* 1960; Beltran-Lopez *et al* 1961; Drever 1961; Ellena *et al* 1987; Chupp *et al* 1989). Therefore $H_{ij}$ is proportional to $g_{ij}$ to certain accuracy. Since a change of $H^{ik}$ to $\lambda H^{ij}$ does not affect $\chi^{ijkl}$ in Eq. (38), we can define $H_{11} = g_{11}$ to remove this scale freedom (Ni 1983a, 1984a).

In Hughes-Drever experiments (Hughes *et al* 1960; Beltran-Lopez *et al* 1961; Drever 1961; Ellena *et al* 1987; Chupp *et al* 1989), $\Delta m/m \leq 0.5 \times 10^{-28}$ or $\Delta m/m_{e.m.} \leq 0.3 \times 10^{-24}$ where $m_{e.m.}$ is the electromagnetic binding energy. Using Eq. (38) in Eq. (31), we have three kinds of contributions to $\Delta m/m_{e.m.}$. These three kinds are of the order of (i) ($H_{\mu\nu}$ - $g_{\mu\nu}$), (ii) ($H_{0\mu}$ - $g_{0\mu}$)v, and (iii) ($H_{00}$ - $g_{00}$)v$^2$ respectively (Ni 1983a, 1984a, 1983b). Here the Greek indices $\mu$, $\nu$ denote space indices. Considering the motion of laboratories from earth rotation, in the solar system and in our galaxy, we can set limits on various components of ($H_{ij}$ - $g_{ij}$) from Hughes-Drever experiments as follows:

$$|H_{\mu\nu} - g_{\mu\nu}| / U \leq 10^{-18},$$
$$|H_{0\mu} - g_{0\mu}| / U \leq 10^{-13} - 10^{-14},$$
$$|H_{00} - g_{00}| / U \leq 10^{-10}, \qquad (39)$$

where $U$ (~ $10^{-6}$) is the galactic gravitational potential.

Eötvös-Dicke experiments (Eötvös 1890; Eötvös *et al* 1922; Roll *et al* 1964; Braginsky and Panov 1971; Schlamminger *et al* 2008 and references therein) are performed on unpolarized test bodies. In essence, these experiments show that unpolarized electric and magnetic energies follow the same trajectories as other forms of energy to certain accuracy. The constraints on Eq. (38) are

$$|1 - \psi| / U < 10^{-10}, \qquad (40)$$

and

$$|H_{00} - g_{00}| / U < 10^{-6}, \qquad (41)$$

where $U$ is the solar gravitational potential at the earth.

In 1976, Vessot and Levine (1979) used an atomic hydrogen maser clock in a space probe to test and confirm the metric gravitational redshift to an accuracy of $1.4 \times 10^{-4}$ (Vessot *et al* 1980), i. e.,

$$|H_{00} - g_{00}| / U \leq 1.4 \times 10^{-4}, \qquad (42)$$

where $U$ is the change of earth gravitational field that the maser clock experienced.

The constraint (40) on the dilaton $\psi$ is stringent. However, with an appropriate mass or potential, the interaction range for dilaton (Cho and Kim 2009; and references therein) or chameleon (Gies *et al* 2008; for recent laboratory experiments on direct detection, see Chen et al 2007 and references therein) becomes intermediate and the associated constraint (40) becomes mild because the corresponding interaction becomes smaller.

With constraints from (i) no birefringence, (ii) no extra physical metric, (iii) no $\psi$ ('dilaton'), we arrive at the theory (31) with $\chi^{ijkl}$ given by (32), i. e., an axion theory (Ni 1983a, 1984a; Hehl and Obukhov 2008). The current constraints on $\varphi$ from astrophysical observations and CMB polarization observations will be discussed in section 5.

From above, we see that for the constraint on $|H_{00} - g_{00}|/U$, Hughes-Drever experiments give the most stringent limit. However, STEP mission concept (Overduin *et al* 2009) proposes to improve the WEP experiment by five orders of magnitude.

This will again lead in precision in determining $H_{00}$.

The theory (31) with $\chi^{ijkl}$ given by (32) is studied in (Ni 1973) and (Ni 1983c). In (31), particles have charges but no spin. To include spin-1/2 particles, we can add the Lagrangian for Dirac particles. One example is given in subsection 2.3. In the next section, we review the experimental tests of the equivalence principle for polarized-bodies.

In the above discussions, we assume $\chi^{(0)ijkl}$ in (33) to be given by the special relativistic value (34). In general, $\chi^{(0)ijkl}$ is determined from cosmological model in a particular theory and provides a framework to test special relativity. From null birefringence of pulsar observations, $\chi^{(0)ijkl}$ is constrained to have the value given in (38) to a precision of $10^{-16}$. From the polarization measurements of extragalactic radio sources, the agreement to special relativity is to $10^{-20}$. With general $\chi^{(0)ijkl}$, instead of general $\chi^{ijkl}$, in the Lagrangian (31), we have a general framework test special relativity in the electromagnetic sector. Its relation with respect to the Mansouri-Sexl framework (Mansouri and Sexl 1977a,b,c) and the Tourrenc-Melliti-Bosredon framework (Tourrenc et al 1996) needs to be studied. Recent works to modify the framework of special relativity include very special relativity (Cohen and Glashow 2006), special relativity triple (Guo et al 2008, 2009), extensions of the Maxwell equations (Lämmerzahl 2005), and Standard Model Extension [SME] (Kostelecky and Mews 2002). SME is a comprehensive framework whose photon sector is the same as $\chi$-g framework with constant $\chi$ and whose fermion sector is a generalization of Lämmerzahl framework (Lämmerzahl 1996, 1998) for Dirac electrons.

To include QCD and other gauge interactions, we have generalized the $\chi$-g framework to general gauge fields (Ni 1987). Lämmerzahl (1996, 1998) formulated a framework for Dirac electrons. A more comprehensive generalization to include a framework to test special relativity, and a framework to test the gravitational interactions of scalar particles and particles with spins together with gauge fields would be ready for exploration. The relation of this generalized framework with respect to the Mansouri-Sexl framework (Mansouri and Sexl 1977a,b,c), the Tourrenc-Melliti-Bosredon framework (Tourrence et al 1996) and Lämmerzahl framework (Lämmerzahl 1996, 1998) is under study.

Theoretically, since a spin 1/2 particle is the most fundamental object in the consideration of quantum spin, one looks into its inertia effects (Hehl and Ni 1990) and curvature effects of a Dirac particle in the standard theory of gravitation as extended by Cartan (1922, 1923), Sciama (1962), Kibble (1961), and Hehl, von der Heyde et al (1976). The inertia effects include the Bonse-Wroblewski phase shift due to acceleration, the Sagnac-type effect, the rotation-spin effect, the redshift of the kinetic energy, and the inertial spin-orbit coupling (Hehl and Ni 1990). The torsion effects are analyzed in Lämmerzahl (1997), Singh and Ryder (1997) and references therein. The curvature effects give the gyrogravitational ratio.

The study of the problem of Dirac particle in gravity from more basic points of view involves two approaches: The first approach starts with study of inertial effects and look into inspirations of novel interaction forms. The second approach starts from an extended framework for Dirac particle in a gravitational field and explores various possible relations between spin and gravity (Lämmerzahl 1996, 1998). In the following, we look into novel interaction forms.

*Spin, Equivalence Principle and Long-Range Forces*

The equivalence principle is an important cornerstone of universal gravitation. The precision of its validity puts an important constraint on gravitation theories and particle theories. Possible deviation from equivalence would give a clue to the microscopic origin of gravity or some new fundamental forces(s). In relation to spins, we look into polarization-dependent deviations from equivalence. In this respect, experiments with polarized entities play an especially important role.

Particles and fields transform appropriately under inhomogeneous Lorentz transformations which form the Poincaré group. The only invariants characterizing irreducible representations of the Poincaré group are mass and spin (or helicity in the case of zero mass). Gravitational interaction is a long-range mass-mass interaction. From a phenomenological approach, we ask whether there is a long-range (or semi-long-range) spin-mass or spin-spin interaction and what is its strength and form of interaction. Experiments on macroscopic spin-polarized bodies are sensitive tools to detect and study these possible interactions to a good precision.

Gauging a subgroup of Lorentz group, Naik and Pradhan (1981) introduce a massless axial vector gauge field, axial photon which gives rise to a super-weak long-range spin-spin force between two particles:

$$V(\underline{r}) = - (g^2/2r)\{ (\underline{S}_1 \cdot \underline{S}_2) + [(\underline{S}_1 \cdot \underline{r})(\underline{S}_2 \cdot \underline{r})/r^2]\}, \tag{43}$$

where $g$ is a coupling constant and $\underline{S}_1$, $\underline{S}_2$ are the spins of two particles. From a precise experiment (Chui and Ni 1993), the strength of anomalous spin-spin interaction is constrained to $(1.2 \pm 2.0) \times 10^{-14}$ of the magnetic spin-spin interaction, and this limits the coupling of axial photon to a level much lower than originally proposed.

From dimensional argument, a spin-spin interaction can have the following form:

$$V(\underline{r}) = (g^2/2r)\{K_1 (\underline{S}_1 \cdot \underline{S}_2) + K_2 [(\underline{S}_1 \cdot \underline{r})(\underline{S}_2 \cdot \underline{r})/r^2]\}, \tag{44}$$

where $K_1$ and $K_2$ are constants. In cosmology, there is a preferred frame. However, in general relativity, there is no specific preferred frame. From this consideration, modern cosmological observations and field theory development, Arkani-Hamed, Cheng, Luty and Mukohyama (2004) proposed an effective field theory of gravity with ghost condensation which gives infrared modification of gravity and has impirically testable predictions. If the standard model fields have direct coupling with the ghost sector, there are spin couplings. Between particle with spins, there is an interaction of the form

$$V(\underline{r}) \sim (M^4/M'^2F^2)(1/r)\{ (\underline{S}_1 \cdot \underline{S}_2) - 3 [(\underline{S}_1 \cdot \underline{r})(\underline{S}_2 \cdot \underline{r})/r^2] \}, \qquad (45)$$

where $M^4/M'^2F^2$ is the coupling constant. (45) has the form of (44) with $K_1 = 1$ and $K_2 = -3$. This gives an effective field theory link to the phenomenological Lagrangian (44). From the experiment of Chui and Ni (1993), the coupling constant $M^4/M'^2F^2$ is constrained to be less than $\sim 10^{-42}$. The effective field theory of Arkani-Hamed *et al* (2004) can induce long-range oscillatory behavior like that in the MOND (Modified Newtonian Dynamics) (Sanders and McGaugh 2002), and therefore can be tested by precision solar-dynamics missions like ASTROD I (Appouchaux *et al* 2009). This theory can also give a cosmic-spin coupling term (Cheng et al 2006) which is the focus of discussion in subsection 4.5.

In the new general relativity of Hayashi and Shirafuji (1979), the coupling with an antisymmetric field leads to a universal spin-spin interaction. From gauging a sub-group of the Lorentz group, Naik and Pradhan (1981) proposed a similar interaction. Around 1980, the particle physics community began to realize the possible existence of Goldstone bosons and/or pseudo-Goldstone bosons. These bosons generate (semi-)long-range forces of monopole-monopole type, monopole-dipole (spin) type, and dipole-dipole (spin-spin) type, just like the new general relativity of Hayashi and Shirafuji (1979). Axion (Weinberg 1987, Wilczek 1978, Dine *et al* 1981, Shifman *et al* 1980, Kim 1979, Cheng *et al* 1995) is such a pseudo-Goldstone boson. The issue of the fifth force arises from the existence of a semi-long-range coupling to baryon number/hyper-charge/lepton number. Attempts have been made to construct models of long-range forces in higher dimensional Kaluza-Klein type theories and superstring theories. All the above cases can be explored experimentally by gravitation-type experiments on macroscopic bodies--Eötvös-type experiments, Galileo-type ("free-fall") experiments and cavendish-type experiments. We will discuss more cases in the next section on lab searches.

*3.5. Gyrogravitational ratio*

Ten years after the discovery of general relativity, Uhlenbeck and Goudsmit (1925, 1926) introduced our present concept of electron spin. From the very beginning of its discovery, spin remains a microscopic object. One way to incorporate spin into the classical general relativity is to treat the aggregate of spins as ordinary angular momentum. This is a standard way to extend general relativity.

However, as we know, for the electromagnetic interaction, the gyromagnetic ratios of elementary particles are different from one, and these ratios reveal the inner electromagnetic structures of elementary particles. Dirac particle has gyromagnetic ratio 2. The gyromagnetic ratio of electron is close to 2; the small deviation is caused by vacuum polarization including QED effect, hadronic effect and others. The gyromagnetic ratio of proton is 5.585 which is totally different from 1 or 2 and reveals electromagnetic structure of proton; neutron is neutron and yet has a large magnetic moment..

For gravitational interaction, we can define gyrogravitational factor as the gravitomagnetic moment divided by angular momentum. The gyrogravitational ratio then normalize that for ordinary angular momentum to be 1. What would be the gyrogravitatinal ratios of elementary particles. If they differ from one, they will definitely reveal some inner gravitational structures of elementary particles. These will give clues to the microscopic origin of gravity.

The Stanford Orbiting Gyro Relativity (GP-B) experiment was launched in 2004 (Everitt *et al* 2008). It aimed at detecting the frame-dragging effect on a gyro. As we have seen in section 3.1., this experiment gives the best constraints on WEP II for nonpolarized bodies. Analysis of GP-B results finds that a nonzero West-East gyroscope drift establishes the frame-dragging effect with a present statistical uncertainty of 15 %. The North-South drift component provides better than 0.5 % measurement of the geodetic effect (Everitt *et al* 2008). It verifies the frame-dragging effect on gyro to about 30 % accuracy with present analysis. *Would intrinsic spin has the same property?* This could be tested by using spin-polarized bodies (e.g. polarized solid $He^3$) instead of rotating gyros in a GP-B type experiment to measure the $He^3$ gyrogravitational ratio (Ni 1983c). Atom interferometry (Berman 1997, Dimopoulos *et al* 2008), nuclear spin gyroscopy (Kornack *et al* 2005) and superfluid $He^3$ gyrometry (Mukharsky *et al* 1999, Avenel *et al* 2004, Chui and Penanen 2005), when developed, may contribute to this very difficult task too.

The measurement of gyrogravitational ratio of elementary particles would probe the microscopic origin of gravity. More specifically, it would probe the following

things:

(i) WEP II for polarized bodies;

(ii) torsion coupling;

(iii) metric-affine connection theory of gravity;

(iv) Yang's approach to gravity;

(v) "The origin of equivalence is identity." conjecture;

(vi) microscopic theories to come in the future.

## 4. Laboratory searches

In this section, we review and discuss electron spin-coupling experiments to search for anomalous interactions --- weak equivalence principle experiments, finite-range spin-coupling experiments, spin-spin coupling experiments and cosmic spin-coupling experiments. An important issue is to make a spin-polarized body. In the next subsection, we describe the strategies to make polarized bodies and methods of measurement for spin-coupling experiment.

*4.1. Polarized bodies and methods of spin-coupling measurement*

To make a polarized-body with a net spin but without net magnetic moment, we need both the orbital angular momentum contribution and spin contribution of magnetic moment so that these contributions cancel each other, with a net total spin remaining.

For iron-group transition elements in a crystal, the elements are exposed to a non-central electric field. In a non-central field the plane of the orbit will move about; the angular momentum components are no longer constant and may average to zero (quenched). For iron-group elements the orbital magnetic moments are mostly quenched although spins may drag some orbital momentum along with them.

For rare-earth elements the orbital angular momentum in the unfilled 4f shell is not quenched. Therefore, rare-earth Fe, Co, or Ni compounds would be good materials for making spin-polarized bodies.

The experiments would be much easier to do at room temperature than at low temperature. For light rare-earth elements and their compounds, the Curie temperatures or Néel temperatures are quite low. Heavy rare-earth compounds, such as Tb, Dy, Ho compounds generally have much higher Curie temperatures. $Dy_6Fe_{23}$ has compensation temperature near room temperature.

In the Dy-Fe compounds, magnetization-vs.-temperature curves indicate anti-ferromagnetic interaction between iron and dysprosium atoms. From the susceptibility data, the strengths of Fe-Fe, Dy-Fe and Dy-Dy exchange interactions can be derived. Fe-Fe exchange interaction dominates the others. Dy-Fe compounds are ferrimagnetic at room temperature. The effective orderings of the iron lattice and dysprosium lattice have different temperature dependence because the strengths of exchange interactions are different. At the compensation temperature, the magnetic moments of two lattices compensate each other so that there is no net magnetization. $Dy^{+++}$ has $L = 5$ and $S = 5/2$. Near this temperature, about half of the Dy magnetization comes from orbital angular momentum, the other half from spin. Most of the iron magnetization comes from spin. So there is a net spin (and net total angular momentum) remaining.

To make polarized bodies, $Dy_6Fe_{23}$ was first synthesized by melting stoichiometric quantities of metallic iron and metallic dysprosium. The $Dy_6Fe_{23}$ ingots were then crushed, pressed into a cylindrical aluminum cup, and magnetized along the desired direction. The magnetic field was shielded by two halves of pure iron casing, a thin aluminum spacer, and two sets of two fitting $\mu$-metal cups with another thin aluminum spacer between the two sets.

Measuring the magnetization-temperature curve of our sample $Dy_6Fe_{23}$, comparing with temperature dependence curve in the literature and calculating from the magnetic properties of $Dy_6Fe_{23}$, there is at least 0.4 net polarized electrons per atom. (Ni 1986)

In addition to the $Dy_6Fe_{23}$ polarized test bodies, we have also made $DyFe_3$, $HoFe_3$, $Ho_6Fe_{23}$ and $Tb_6Fe_{23}$ samples.

Graham and Newman (1983, 1986; Graham 1987) used carefully prepared hybrid split rings of $GdNi_5$/NdNi at superconducting temperatures as spin polarized bodies both for spin sources and spin detectors in their experiment to search for anomalous (nonmagnetic) spin-spin interactions. The split ring is made of two halves: one half is made of $GdNi_5$ whose spins are parallel to J (angular momentum) and hence antiparallel to B while the other half is made of NdNi whose spins are antiparallel to J and hence parallel to B. When the ring is magnetized with a toroidal winding the resulting B field lines are continuous around the ring and largely contained within it, and yet the two halves of the ring have spin alignments in the same direction, providing a large net spin alignment in ring. The split ring is enclosed in superconducting shields and magnetization is done after the shields have been cooled to superconducting temperature.

In the cosmic-spin coupling experiment of Heckel *et al* (2006, 2008), their polarized spin pendulum is constructed from 4 octogonal "pucks", i.e., split rings, one side of puck is AlNiCo and the other side is $SmCo_5$. AlNiCo is a conventional soft ferromagnet in which the magnetic field is created almost entirely by electron spins. $SmCo_5$ is a hard rare-earth magnet in which the orbital magnetic moment of the electrons in the $Sm^{3+}$ ion nearly cancels their spin magnetic moment. After each "puck" was assembled, the AlNi Co was magnetized to the same degree as the $SmCo_5$ by sending appropriate current pulses through coils temporarily wound around the pucks. Four such pucks were stacked and magnetically shielded to make the polarized spin pendulum.

To design and make a polarized body, there are two factors to consider: first, the net spin density should be large; second the polarized body should be magnetically shielded as much as possible in order to detect and measure anomalous (nonmagnetic) interactions. For the first, we need permanent magnetic materials. For the second, we need soft magnetic materials or superconducting shields. When the basic principles of magnetism are acquainted, one would be able to design and make one's own spin-polarized bodies.

Once spin-polarized bodies are made, we can talk about methods of measurement. For the spin-coupling experiment, the interaction is between two bodies. One body is considered to be the source. The coupling or effect on the other is to be sensed, and therefore this other body is the beginning of the sensing chain. The force or the induced spin effect is to be picked up by a transducer which could be a torsion pendulum with an optical sensing level or a SQUID transducer with a pickup circuit. To search for anomalous spin-spin interactions, the first polarized body needs to have a good magnetic shielding and its motion modulated in order to differentiate the signal from surrounding background noise. Anomalous spin interactions would induce motion of electrons and this motion will induce magnetic field change. The detection mechanism could be polarized body (bodies) on a torsion balance or paramagnetic salt with a SQUID transducer.

To detect the spin-mass interactions there are two kinds of method. The first kind is to detect force on the mass. A torsion balance with un-polarized bodies will do. The second kind is to detect force on the spin. Either a torsion balance with polarized bodies (body) or paramagnetic salt with a SQUID transducer will do.

For searching cosmic-spin interactions, since cosmos has little net spin polarization, we need either a torsion balance with polarized body (bodies) or paramagnetic salt with a SQUID transducer.

Regarding to the properties of torsion balance, we refer the readers to Gillies and Ritter (1993), Kuroda (1995) and Ritter *et al* (1999) for references.

*4.2. The weak equivalence principle experiments*

In this section, we discuss weak equivalence principle experiments using polarized bodies, and GP-B experiment as a weak equivalence principle experiment with rotating bodies. A comprehensive compilation on weak equivalence experiments (polarized and unpolarized) before 1997 can be found in Gillies (1987, 1991, 1997).

Before mid-seventies, the actual weak equivalence experiments were performed on unpolarized bodies. These experiments constrained only 2 degrees of freedom of χ. Only when experiments are performed on polarized bodies with various electromagnetic energy configurations, can they constrain the other 18 degrees of freedom. This situation motivated many people to study other existing and potential experimental and observational evidences for EEP, and to perform experiments on polarized bodies and to search for spin-dependent forces.

*4.2.1. Polarized equivalence principle experiments*
To investigate the equivalence principle for spin-polarized bodies or to probe the mass-spin (monopole-spin/baryon-number-to-spin) interactions, we have used both a beam balance (Ni *et al* 1988; Hsieh *et al* 1989; Ni *et al* 1990) and a torsion balance (Chou *et al* 1990) to test a magnetically shielded spin-polarized body of $Dy_6Fe_{23}$. From these results, we have inferred that, to an accuracy of $5 \times 10^{-3}$, the polarized electron falls at the same rate as unpolarized bodies in the earth's gravitational field (Ni *et al* 1990), and that it falls at the same rate as unpolarized bodies in the solar gravitational field with a deviation from this unity ratio estimated at $(3 \pm 4) \times 10^{-2}$ (Chou *et al* 1990).

In 2001, we have used a rotatable torsion-balance experiment and have improved the solar equivalence principle test by one order of magnitude (Hou and Ni 2001). A schematic experimental setup is shown in Figure 1; the pan-set configuration is shown on the right of the figure. Our spin-polarized body contains 23.81 g of $HoFe_3$. the equivalence with respect to unpolarized aluminum-brass masses is $\eta^S_0 = (-0.68 \pm 0.9) \times 10^{-9}$, $\eta^S_p = (1.8 \pm 5.3) \times 10^{-9}$ in the solar gravitational field and $\eta^E_0 = (-0.24 \pm 0.55) \times 10^{-9}$ in the earth gravitational field. This result indicates that to $(2.3 \pm 6.6) \times 10^{-3}$ the polarized electron falls with the same rate as unpolarized bodies in the solar gravitational field; and that to $(2.6 \pm 7.5) \times 10^{-5}$ the polarized nuclei falls with the same rate as unpolarized bodies in the solar gravitational field. This improves our previous result by one order of magnitude for polarized electron and by about 50 times for polarized nuclei for the solar gravitational field. Dedicated low-temperature experiments for testing the weak equivalence principles for polarized nucleus have been proposed and under implementation (Daniels and Ni 1991, 1994 and references therein). Atom interferometry for atoms or molecules with polarized nuclei will be good for polarized-nucleus equivalence principle tests also.

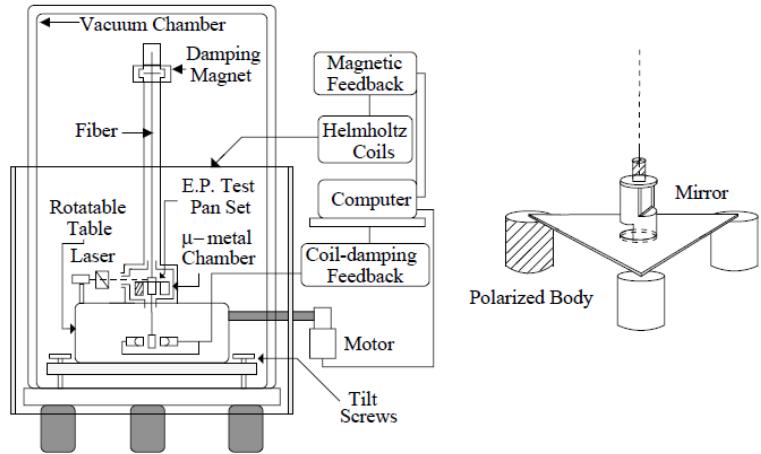

Fig. 1. Schematic experimental setup with the pan-set configuration shown on the right of the diagram.

*4.2.2. GP-B experiment as a WEP II experiment*

The aim of Gravity Probe-B Relativity Gyroscope Experiment (GP-B) experiment is to measure the frame dragging effect in general relativity (GR). It carries 4 gyroscopes (quartz balls) pointing to the guide star IM Pegasi in a polar orbit of height 642 km (Figure 2). The GR prediction of frame-dragging effect and geodetic effect is to be measured. The prediction is shown in Figure 2, and compiled together with the solar geodetic and proper motion effects. (Everitt *et al* 2008, Heifetz *et al* 2008, Keiser *et al* 2008, Muhlfelder *et al* 2008, Silbergleit *et al* 2008)

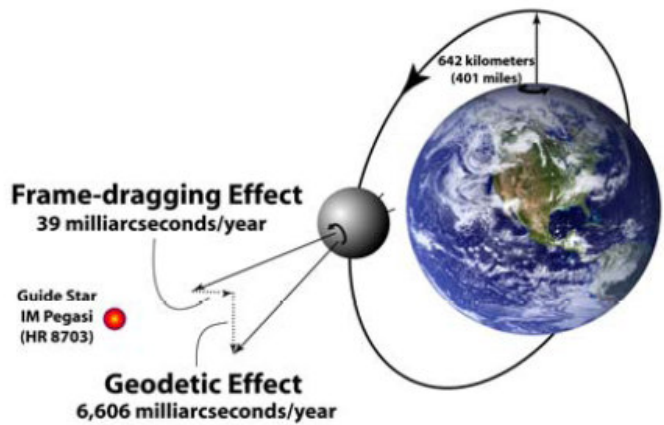

**Figure 2.** Predicted precessions of the GP-B gyroscope (Everitt *et al* 2008) (North and West are positive in this coordinate system)

**Table 5.** Compilation of NS (north-south) and EW (east-west) components of earth effect, solar geodetic effect and proper motion effect on GP-B gyros. (Everitt *et al* 2008)

|     | Earth | Solar Geodetic | Proper Motion | Net Expected |
|-----|-------|----------------|---------------|--------------|
| NS  | -6606 | +7             | +28±1         | -6571±1      |
| EW  | -39   | -16            | -20±1         | -75±1        |

GP-B experiment measured the frame-dragging precession of a gyroscope in a polar orbit of height 642 km to within 30% of the General Relativity value 39 mas/year in their ongoing analysis (Figure 3). This means that the rotation state of the quartz gyros with different rotational speeds agree with that of local inertial frame to about 12 mas/year. In torsion theories like New General Relativity of Hayashi and Shirafujii (1979), torsion can be generated by macroscopic rotating and affect macroscopic rotating body. They (Hayashi and Shirafujii 1990) find that New General Relativity gives 1+ $x$ times the General relativity value of frame-dragging (see also Mao *et al* 2007, Flanagan and Rosenthal 2007). Their $x$ is parameter arisen from a new quantum occurring in a spin-spin interaction of two fermions. They also infer on a theoretical reason that $x$ is unity. GP-B gives a limit on $x$ to be $|x| < 0.3$.

Although GP-B has not fully achieved its goal yet, it is successful in verifying the WEP II for unpolarized body down to the curvature level. Together with the measurement of the spin inertial effect, it serves as a starting point for the measurement of gyrogravitational factor of particles.

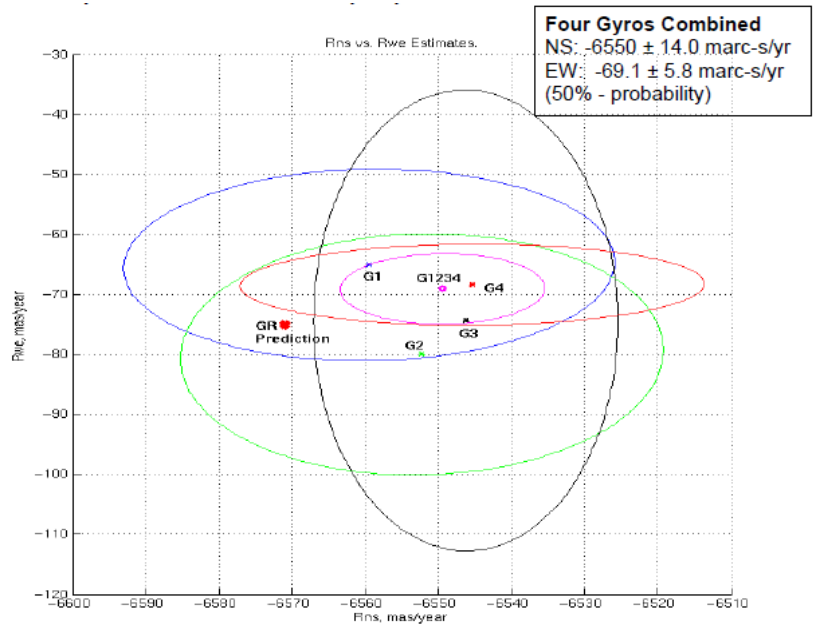

**Figure 3.** Preliminary Relativity Estimates. Gyroscopes 1, 2, 4: Segments 5, 6,

and 9, (Gyro 3: Segments 5, 6), not including systematic error or a model sensitivity analysis. (Everitt *et al* 2008)

*4.3. Finite-Range Spin-Coupling Experiments searching for anomalous spin-monopole couplings*

In Jen *et al* (1992), we used torsion balance with two cylindrical copper test masses and two cylindrical polarized "attracting" $Dy_6Fe_{23}$ masses to search for finite-range mass-spin interactions with the Hamiltonian of the form $H_{int} = f(r)\sigma \cdot \hat{\mathbf{r}}$. This result showed that for the range of 3-5 cm, the upper limit of this interaction for our test mass and the $Dy_6Fe_{23}$ polarized mass was below 1% of their gravitational interaction. We considered, in particular, the case of $f(r) = -Ae^{-\mu r}mU$ with $U$ the gravitational potential of the unpolarized body; that is, the finite range mass-spin interaction is of the following form

$$H_{int} = -Ae^{-\mu r}mU\sigma \cdot \hat{\mathbf{r}}. \tag{46}$$

Ritter, Winkler and Gillies (1993) use spin-polarized $Dy_6Fe_{23}$ masses acting on unpolarized copper masses in a dynamic-mode torsion pendulum, and searched for interaction of the axion (Weinberg 1978, Wilczek 1978, Kim 1979, Dine *et al* 1981, Shifman *et al* 1980, Cheng *et al* 1995; Moody and Wilczek 1984) form

$$H_{int} = [\hbar(g_s g_p)/8\pi mc] (1/\lambda r + 1/r^2)exp(-r/\lambda)\, \sigma \cdot \hat{\mathbf{r}}. \tag{47}$$

In (47), $\lambda$ is the range of the interaction, $g_s$ and $g_p$ are the coupling constants of vertices at the polarized and unpolarized particles and m is the mass of the polarized particle. Constraints on the coupling $g_s g_p/\hbar c$ with respect to the range from various experiments are plotted in a logarithmic plot (Figure 4). For $\lambda < 0.3$ m, Jen *et al* (1992) and Ritter *et al* (1993) give more stringent constraints than Wineland *et al* (1991) and Venema *et al* (1992), and for $\lambda > 0.3$ m, vice versa. Wineland *et al* (1991) and Venema *et al* (1992) investigate the existence of hypothetical anomalous spin-dependent forces by sensing the interaction of polarized trapped ions with fermions in the earth. These experiments are more sensitive to longer range forces, while experiments with laboratory sources are more sensitive to shorter range forces. Youdin *et al* (1996) has the best limit for 0.1 m $< \lambda <$ 8 m.  Our works (Ni *et al* 1999; Jen *et al* 1992) have the best limit for $\lambda < 0.1$ m. In figure 4, we also show the proposed sensitivities of the STEP spin-coupling experiment (Shaul *et al* 1996; ESA SCI 1993) and the AXEL spin-coupling experiment (Ni 1996; Li and Ni 2000; Ni 2000) together with allowed region of present axion models.

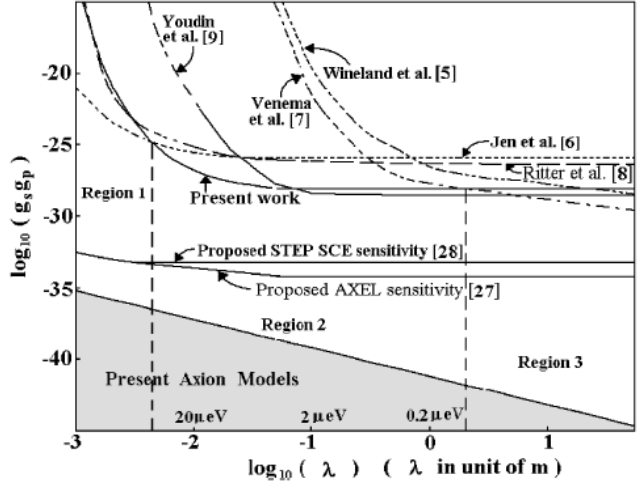

Figure 4. Limits on **σ·r** spin coupling for axionlike interactions
from various experiments.

Recently, there is a measurement in the shorter range. Hammond *et al* (2007, 2008) has developed a new superconducting torsion balance to detect force on the mass for the spin-coupling experiment. This experiment has the best sensitivity in the shorter range as in Fig. 5. For comparison, the gravitational interaction between an electron and a nucleon, separated by $\lambda$ is also shown.

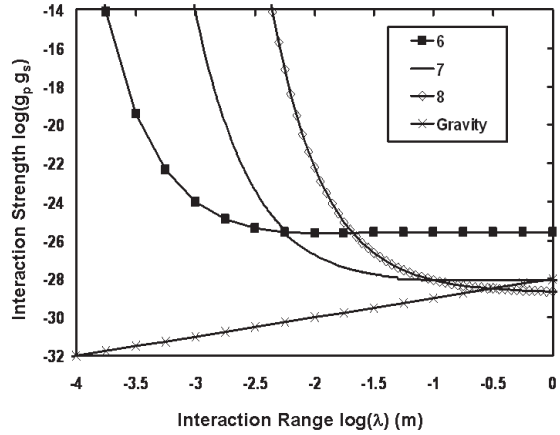

Figure 5. Current experimental limits on pseudoscalar in the short range
couplings as a function of interaction range. (Hammond *et al* 2008)

In Fig. 4, there are magnetic resonance experiments, torsion balance experiments and SQUID experiments. In the following, we give a taste of experimental procedure using a SQUID experiment (Ni *et al* 1999). The experiment measures the effective **B**$_{eff}$ field produced by hypothetical axion or axion-like interaction while magnetic field is shielded by two niobium superconducting shields. Equation (47) can be written in the form

$$H_{int} = -\mathbf{m} \cdot \mathbf{B}_{eff} = -\mu_e \boldsymbol{\sigma} \cdot \mathbf{B}, \tag{48}$$

with $\mu_e = -|\mu_e|$ the magnetic moment of the electron. Hence in this experiment, we sensitivly measure $\mathbf{B}_{eff}$ field given by

$$\mathbf{B}_{eff} = -\frac{\hbar}{\mu_e} \frac{g_s g_p}{8\pi m_e c} \left(\frac{1}{\lambda_r} + \frac{1}{r^2}\right) \exp(-r/\lambda) \hat{\mathbf{r}}. \tag{49}$$

The scheme of our experimental setup is shown in Figure 6. Our copper mass is sitting on one side of the turntable underneath the dewar. In the data taking, a laser beam and a chopper-photodetector system is used to lock the output signal of the dc SQUID to the rotation angle of the polarized bodies. The laser beam is intercepted by the chopper when the copper axis is in line with the axis of the paramagnetic salt. We define this angle to be zero degree, and expect the $\boldsymbol{\sigma} \cdot \mathbf{r}$ interaction signal to be proportional to cos θ, where θ is the angular position of the copper mass.

To start the measurement, we set the turntable with copper mass rotating at 0.96 cycle per second with a stepping motor system. The stability of the rotation speed is better than $10^{-4}$. The output of voltage of the dc SQUID system for $1\phi_o$ from the most sensitive scale of the dc SQUID controller is 10 V. This output is further amplified 1000 times and low-pass filtered to 10 Hz bandwidth, and then read into a computer with an analog to digital converter. The angular position of the copper mass is simultaneously read into this computer. The typical noise of the SQUID output after 1000 times amplification and 10 Hz low pass filtering as recorded by ADC (Analog-to-Digital Converter) is about ± 300 mV. This is consistent with the dc SQUID noise 200 mV/$\sqrt{Hz}$ after amplification. When we average the data for 400 cycles, the typical output is about ± 50 mV and the pattern repeats. And this pattern sustains after we take away the copper mass. Hence, we infer that this pattern is largely interference background (due to electronics, electric couplings, etc.) To subtract this interference background, we average the data for 4-5 hours, alternatively take away and put back the copper cylinder to average the data for another 4-5 hours and subtract the results to find the net effects.

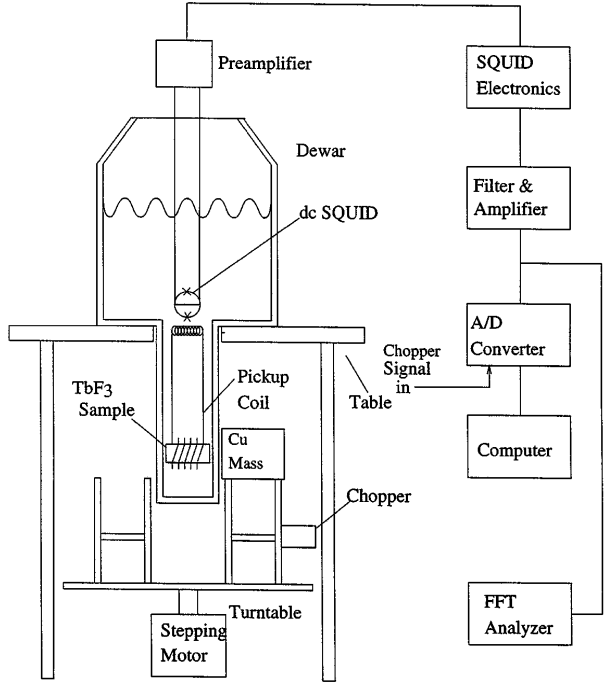

Fig. 6. Schematic for spin-coupling experiment.

The weighted average of the six runs for the amplitude of $\cos\theta$ component is $(0.49 \pm 2.34)$ mV. Expressed in terms of flux amplitude, it becomes $(0.49 \pm 2.34) \times 10^{-7}$ $\phi_o$. Converted to $\mathbf{B}_{eff}$, we have $(1.13 \pm 5.38) \times 10^{-12}$ Gauss and the coupling constant $g_s g_p/\hbar c$ is $(0.14 \pm 0.67) \times 10^{-28}$ for $\lambda > 30$ mm. Our experimental constraint on the coupling constant $g_s g_p/\hbar c$ improves over previous results by 2 orders of magnitude at $\lambda = 30$ mm. Further improvement will be implemented. For finite-range Leitner-Okubo-Hari Dass interaction, the dimensionless parameter A is constrained to less than 10 for the range parameter $\lambda = \mu^{-1} > 30$ mm.

*4.4. Experiments searching for anomalous spin-spin interactions*

Usually the dipole-monopole (spin-mass) part of an interaction is larger than its dipole-dipole (spin-spin) part. Monopole-monopole part is sometimes larger if there is no constraint. However the monopole-monopole part does not change with polarization and is usually harder to detect. Therefore in search for a new interaction, searching for dipole-monopole part is usually more significant. This is true for axion search. However, for axial photon (Naik and Pradhan 1981; Pradhan *et al* 1985) and arion (Ansel'm and Ural'tsev 1982; Ansel'm 1982) search, the search for anomalous spin-spin interactions give stringent constraints.

Let $\alpha_s$ be the strength of the anomalous spin-spin interaction compared to the magnetic spin-spin interaction. The pioneer work of Graham and Newman (1986; Graham 1987) used carefully prepared hybrid split toroids of GdNi$_5$/NdNi at superconducting temperatures with a torsion balance of the feedback deflection type. Their experimental result assigns uncertainties in two parts: statistical at the 1σ level and an estimated systematic uncertainty $α_s$ = (8.0 ± 6.3 ± 1.1) × $10^{-11}$. The torsion balance experiment led by Ritter at the University of Virginia uses the period method and gives the constraint $α_s$ = (1.6 ± 6.9) × $10^{-12}$ (Ritter *et al* 1990). Using the deflection method with the torsion pendulum, Pan *et al* (1992) improves this constraint to $|α_s|$ < 1.5 × $10^{-12}$. Adapting the induced ferromagnetism method of Vorobyov and Gitarts (1988) to paramagnetism, we use a low noise dc SQUID system to search for the interaction of spins in a spin-polarized test mass and those in a paramagnetic salt, separated by a *μ*-metal shield and a double-layer superconducting shield (Chui and Ni 1993; Ni *et al* 1993, 1994). Our results limit the strength of $α_s$ to $α_s$ = (1.2 ± 2.0) × $10^{-14}$ (Ni *et al* 1994). This limits the coupling of axial photon and the arion coupling to a level much lower than originally proposed. This also limits the coupling constant of ghost-condensation effective theory of Arkani *et al* (2004).

*4.5. The cosmic-spin coupling experiments*

Hughes-Drever experiments (Hughes *et al* 1960; Beltran-Lopez 1961; Drever 1962; Ellena et al 1987; Chupp *et al* 1989) test the Cosmic Spatial Isotropy for spin 3/2 particle very precisely. Their constraints in the photon sector and in the electron sector have been discussed in subsection 3.4 and subsection 2.3 respectively. Frequency and clock experiments push this limit even further (Wineland *et al* 1991); the relative constraints are listed in Table 4 in subsection 3.2 and Table 6 in this subsection..

As to the spin 1/2 particle, Phillips (1987) used a cryogenic torsion pendulum carrying a transversely polarized magnet with superconducting shields to set a stringent limit of 8.5 × $10^{-18}$ eV for the splitting of the spin states of an electron at rest on Earth. In our laboratory we have used a room-temperature torsion balance with a magnetically-compensated polarized-body and set a spin energy level splitting limit of 3 × $10^{-18}$ eV (Wang *et al* 1993; Chang *et al* 1995). Berglund *et al* (1995) use a magnetic resonance technique and set a limit of 1.8 × $10^{-18}$ eV on the energy splitting.

For the analysis of cosmic anisotropy for electrons, we can use the following Hamiltonian:

$$H_{cosmic} = C_1σ_1 + C_2σ_2 + C_3σ_3 \qquad (50)$$

in the cosmic frame of reference. This includes the following two cases, (i) $H_{cosmic}$ = $g\sigma \cdot \mathbf{n}$ with $C_1 = g\mathbf{n}_1$, $C_2 = g\mathbf{n}_2$, $C_3 = g\mathbf{n}_3$ as considered in Chen *et al* (1992), Wang *et al* (1993), Chang *et al* (1995); here C's are constants; theories with noncommutative spacetime geometries can also give such a term (Ansimov *et al* 2002, Hinchliffe *et al* 2004), (ii) $H_{cosmic} = g\sigma \cdot \mathbf{v}$ with $C_1 = g\mathbf{v}_1$, $C_2 = g\mathbf{v}_2$, $C_3 = g\mathbf{v}_3$ as considered in the context of Phillips (1965, 1987), Phillips and Woolum (1969), Stodolsky 1975, Nilson and Picek 1983, Hou and Ni 1997); in this case, since $\mathbf{v}$ is largely the velocity of our solar system through the cosmic preferred frame, to a first approximation, C's are also constants; the ghost-condensate theory of Arkani-Hamed *et al* (2004) and Cheng *et al* (2006) can also give such a term. For convenience, we use the celestial equatorial coordinate system from the center of earth for our laboratory position, i.e., the earth rotation axis (North pole direction) as z-axis and the direction of the spring equinox as the positive x-direction. All the above experimental constraints are on $C_1$ and $C_2$. The constraints on $C_3$ are crude.

To improve the precision and to constrain on $C_3$, we used a rotatable torsion balance carrying a transversely spin-polarized ferromagnetic $Dy_6Fe_{23}$ mass after 1996 (Hou and Ni 1997, Hou *et al* 2000, Hou *et al* 2003). With a rotation period of one hour, the period of anisotropy signal is reduced from one sidereal day by about 24 times, and hence the 1/$f$ noise is greatly reduced. Our present experimental results constrain the cosmic anisotropy constant $C_1$, $C_2$, $C_3$ to $\sqrt{C_1^2 + C_2^2} < 3\times 10^{-20}$ eV and | $C_3$| < 7 × 10$^{-19}$ eV. This improved the previous limits on ($C_1$, $C_2$) by 27 times and $C_3$ by a factor of 500 (Hou *et al* 2003).

The angular velocity of the cosmic signals is $\Omega + \omega$, $\Omega - \omega$, and $\omega$. Here $\omega$ is the angular velocity of rotatable table and $\Omega$ is the angular velocity of the earth. By the earth rotation the projection of the electron spin in the x-y plane rotates to opposite direction relative to the neutrino background or cosmos after half of a sidereal day (11 hours 58 min 2 seconds). Adding the two data sets separated by half sidereal day, we can eliminate the $\Omega + \omega$, $\Omega - \omega$ term, and estimate $C_3$. Subtracting between the same two data sets, we can eliminate the $\omega$ term. With 4 sequential data sets (each set's separated by half sidereal day) in opposite rotational direction of rotatable table, the signals with frequencies $\Omega + \omega$, $\Omega - \omega$, $\omega$ can be separated. The results of eight such sets of runs gives the limits on $C_1$, $C_2$, $C_3$ just mentioned. This experiment also gives a stringent CPT test in the SME framework which has the Hamiltonian (50) as part of their framework. Our constraint on the Lorentz and CPT violation parameters $b^e_\perp$ and $b^e_Z$ of Bluhm and Kostelecky (2000) is $b^e_\perp$ [=$(C_1^2+C_2^2)^{1/2}$] $\leq 3.1\times 10^{-29}$ GeV and |$b^e_Z$| [=|$C_3$|] $\leq 7.1\times 10^{-28}$ GeV.

Heckel *et al* (2006, 2008) used a rotatable torsion pendulum with a spin-polarized body described in subsection 4.1 to perform a cosmic-spin coupling experiment. They have the best results to date. As we discussed in subsection 3.2, their results measure the inertial spin-rotation effect to 5 % and verify it to 18% of the special relativity value.

The constraints from all the cosmic-spin coupling experiments using electron spins are compiled in Table 6.

**Table 6.** Cosmic-spin coupling experiments using electron spins. $\delta E_\perp = 2(C^2_1 + C^2_2)^{1/2}$ and $\delta E_\parallel = 2|C_3|$ are the energy level splitting parallel and transverse to the earth rotation axis, respectively.

| Reference | $\delta E_\perp$ ($10^{-18}$) eV | $\delta E_\parallel$ ($10^{-18}$) eV |
|---|---|---|
| Phillips (1987) | ≤ 8.5 | N.A. |
| Wineland *et al* (1991) | ≤ 550 | ≤ 780 |
| Chen *et al* (1992) | ≤ 7.3 | N.A. |
| Wang *et al* (1993) | ≤ 3.9 | N.A. |
| Chang *et al* (1995) | ≤ 3.0 | N.A. |
| Berglund *et al* (1995) | ≤ 1.7 | N.A. |
| Hou *et al* (2003) | ≤ 0.06 | ≤ 1.4 |
| Heckel *et al* (2006) | ≤ 0.0004 | ≤ 0.01 |

**5. Astrophysical and cosmological searches**

*5.1. Constraints from astrophysical observations prior to CMB polarization observations*

In section 3.4., we have reviewed using the pulsar timing observations, radio galaxy observation, and optical polarization observation of cosmologically distant astrophysical sources to constrain the χ-g framework to two metric, one scalar and one pseudoscalar. Hughes-Drever experiments then constrain the two metric to be the same to high accuracy. Eötvös experiments on unpolarized bodies constrain the scalar to be nearly one. Only the pseudoscalar is largely not constrained.

For the gravity theory (21) with an effective pseudoscalar, discussed in section 2.4. and section 3.4., the electromagnetic wave propagation equation is given by equation (23). In a local inertial (Lorentz) frame of the *g*-metric, it is reduced to

$$F^{ik}{}_{,k} + e^{ikml} F_{km} \varphi_{,l} = 0. \tag{51}$$

Analyzing the wave into Fourier components, imposing the radiation gauge condition, and solving the dispersion eigenvalue problem, we obtain $k = \omega + (n^\mu \varphi_{,\mu} + \varphi_{,0})$ for right circularly polarized wave and $k = \omega - (n^\mu \varphi_{,\mu} + \varphi_{,0})$ for left circularly polarized wave in the eikonal approximation (Ni 1973, Carroll *et al* 1990). Here $n^\mu$ is the unit 3-vector in the propagation direction. The group velocity is

$$v_g = \partial\omega/\partial k = 1, \tag{52}$$

independent of polarization. There is no birefringence. For the right circularly polarized electromagnetic wave, the propagation from a point $P_1 = \{x_{(1)}{}^i\} = \{x_{(1)}{}^0; x_{(1)}{}^\mu\} = \{x_{(1)}{}^0, x_{(1)}{}^1, x_{(1)}{}^2, x_{(1)}{}^3\}$ to another point $P_2 = \{x_{(2)}{}^i\} = \{x_{(2)}{}^0; x_{(2)}{}^\mu\} = \{x_{(2)}{}^0, x_{(2)}{}^1, x_{(2)}{}^2, x_{(2)}{}^3\}$ adds a phase of $\alpha = \varphi(P_2) - \varphi(P_1)$ to the wave; for left circularly polarized light, the added phase will be opposite in sign (Ni 1973). Linearly polarized electromagnetic wave is a superposition of circularly polarized waves. Its polarization vector will then rotate by an angle $\alpha$. Locally, the polarization rotation angle can be approximated by

$$\begin{aligned}\alpha = \varphi(P_2) - \varphi(P_1) &= {}_i\Sigma_0^3\,[\varphi_{,i} \times (x_{(2)}{}^i - x_{(1)}{}^i)] = {}_i\Sigma_0^3\,[\varphi_{,i}\Delta x^i] = \varphi_{,0}\Delta x^0 + [{}_\mu\Sigma_1^3 \varphi_{,\mu}\Delta x^\mu] \\ &= {}_i\Sigma_0^3\,[V_i \Delta x^i] = V_0 \Delta x^0 + [{}_\mu\Sigma_1^3 V_\mu \Delta x^\mu]. \end{aligned} \tag{53}$$

The rotation angle in (53) consists of 2 parts -- $\varphi_{,0}\Delta x^0$ and $[{}_\mu\Sigma_1^3 \varphi_{,\mu}\Delta x^\mu]$. For light in a local inertial frame, $|\Delta x^\mu| = |\Delta x^0|$. In Fig. 7, space part of the rotation angle is shown. The amplitude of the space part depends on the direction of the propagation with the tip of magnitude on upper/lower sphere of diameter $|\Delta x^\mu| \times |\varphi_{,\mu}|$. The time part is equal to $\Delta x^0 \varphi_{,0}$. ($\nabla\varphi \equiv [\varphi_{,\mu}]$) When we integrate along light (wave) trajectory in a global situation, the total polarization rotation (relative to no $\varphi$-interaction) is again $\Delta\varphi = \varphi_2 - \varphi_1$ for $\varphi$ is a scalar field where $\varphi_1$ and $\varphi_2$ are the values of the scalar field at the beginning and end of the wave. When the propagation distance is over a large part of our observed universe, we call this phenomenon cosmic polarization rotation (Ni 2008).

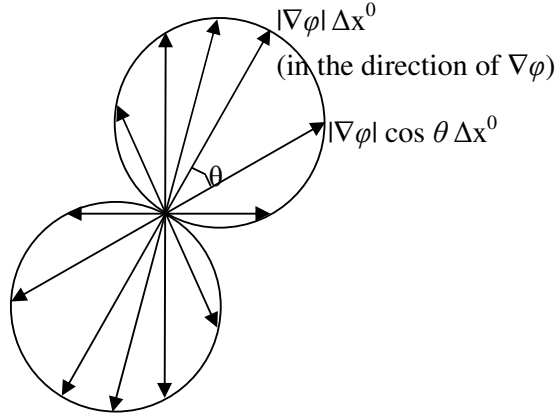

**Figure 7.** Space contribution to the local polarization rotation angle -- $[\sum_{\mu=1}^{3}\varphi_{,\mu}\Delta x^{\mu}] = |\nabla\varphi|\cos\theta\,\Delta x^{0}$. The time contribution is $\varphi_{,0}\,\Delta x^{0}$. The total contribution is $(|\nabla\varphi|\cos\theta + \varphi_{,0})\,\Delta x^{0}$. ($\Delta x^{0} > 0$).

In the CMB polarization observations, there are variations and fluctuations. The variations and fluctuations due to scalar-modified propagation can be expressed as $\delta\varphi(2) - \delta\varphi(1)$, where 1 denotes a point at the last scattering surface in the decoupling epoch and 2 observation point. $\delta\varphi(2)$ is the variation/fluctuation at the last scattering surface. $\delta\varphi(1)$ at the present observation point is zero or fixed. Therefore the covariance of fluctuation $<[\delta\varphi(2) - \delta\varphi(1)]^2>$ gives the covariance of $\delta\varphi^2(2)$ at the last scattering surface. Since our Universe is isotropic to $\sim 10^{-5}$, this covariance is $\sim (\xi \times 10^{-5})^2$ where the parameter $\xi$ depends on various cosmological models. (Ni 2008, 2009a)

For linearly polarized wave, there is an induced rotation of polarization with an angle of $(n^{\mu}\varphi_{,\mu} + \varphi_{,0}) = \Delta\varphi = \varphi_2 - \varphi_1$ where $\varphi_1$ and $\varphi_2$ are the values of the scalar field at the beginning and end of the wave. When we integrate along light (wave) trajectory, the total polarization rotation (relative to no $\varphi$-interaction) is $\Delta\varphi = \varphi_2 - \varphi_1$ where $\varphi_1$ and $\varphi_2$ are the values of the scalar field at the beginning and end of the wave. When the propagation distance is over a large part of our observed universe, we call this phenomenon cosmic polarization rotation.

Now we must say something about nomenclature.

Birefringence, also called double refraction, refers to the two different directions of propagation that a given incident ray can take in a medium, depending on the direction of polarization. The index of refraction depends on the direction of polarization.

Dichroic materials have the property that their absorption constant varies with polarization. When polarized light goes through dichroic material, its polarization is rotated due to difference in absorption in two principal directions of the material for the two polarization components. This phenomenon or property of the medium is called dichroism.

In a medium with optical activity, the direction of a linearly polarized beam will rotate as it propagates through the medium. A medium subjected to magnetic field becomes optically active and the associated polarization rotation is called Faraday rotation.

Cosmic polarization rotation is neither dichroism nor birefringence. It is more like optical activity, with the rotation angle independent of wavelength. Conforming to the common usage in optics, one should not call it cosmic birefringence.

In 1973, I used the laboratory experiments such as Hughes-Drever experiments and atomic-level measurement to constrain the pseudoscalar to about $10^{10}$ and propose to use electromagnetic propagation of astrophysical distances to obtain better constraints in the future (Ni 1973). The electromagnetic propagation of astrophysical distance from pulsars and radio galaxies is then used to constrain the χ-g framework (Ni 1983, 1984a, 1984b), discussed in subsection 3.4., and the polarization rotation angle due to pseudoscalar/constant vector (Carroll *et al* 1990) which will be discussed in the following.

Carroll, Field and Jackiw (1990) used the fact that the distribution of the difference between the position angle of the radius axis and the position angle of the E vector of linear radio polarization in distant radio galaxies, with redshift between 0.4 and 1.5, peaks around 90 deg to argue that this phenomenon is intrinsic to the source and therefore put limits ($\Delta\varphi \leq 6.0°$ at the 95% confidence level) on the rotation of the plane of polarization for radiation travelling over cosmic distances. Cimatti *et al* (1994) and di Serego Alighieri *et al* (1995) used the perpendicularity between the optical/UV axis and the optical/UV linear polarization of distant radio galaxies, as expected since the latter is due to scattering of anisotropic nuclear radiation, to show that the plane of polarization is not rotated by more than 10° for every distant radio galaxy with a polarization measurement up to z = 2.63. The advantage of Cimatti *et al* (1994) using optical polarization is that it is based on a physical prediction of the polarization orientation due to scattering and it does not require Faraday rotation correction (di Serego Alighieri 2006).

In 1997, Nodland and Ralston (1997) announced that they found an additional rotation of synchrotron radiation from distant radio galaxies and quasars which is independent of wavelength. However, other people before the announcement and after the announcement did not find this in their analyses and put a limit of $\Delta\varphi \leq 0.17$-1 (rad) over cosmological distance from polarization observations of radio galaxies (Carroll and Field 1991, Cimatti *et al* 1994, di Serego Alighieri *et al* 1995, Wardle *et al* 1997, Eisenstein and Bunn 1997, Carroll and Field 1997, Carroll 1998, Loredo *et al* 1997). In particular, Cimanti, di Serego Alighieri, Field, and Fosbury had found no rotation within 10 degrees (0.17 rad) for the optical/UV polarization of radio galaxies for all radio galaxies with 0.5<z<2.6 in their list (Cimatti *et al* 1994, di Serego Alighieri *et al* 1995). There is also no rotation within 10 degrees for a 2006 update with z > 2.0 (up to z = 4.1) (di Serego Alighieri 2006). More recent work (di Serego Alighieri S *et al* 2009) confirms this with better constraints.

*5.2. Constraints on cosmic polarization rotation from CMB polarization observation*

Now we review and compile the constraints of various analyses from CMB polarization observations.

In 2002, DASI microwave interferometer observed the polarization of the cosmic background (Kovac *et al* 2002). E-mode polarization is detected with 4.9 σ. The TE correlation of the temperature and E-mode polarization is detected at 95% confidence. This correlation is expected from the Raleigh scattering of radiation. However, with the (pseudo)scalar-photon interaction under discussion, the polarization anisotropy is shifted differently in different directions relative to the temperature anisotropy due to propagation; the correlation will then be downgraded. In 2003, from the first-year data (WMAP1), WMAP found that the polarization and temperature are correlated to more than 10 σ (Bennett *et al* 2003). This gives a constraint of about $10^{-1}$ for $\Delta\varphi$ (Ni 2005a,b).

Further results and analyses of CMB polarization observations came out after 2006. In Table 7, we update our previous compilations (Ni 2008, 2009a). Although these results look different at 1 σ level, they are all consistent with null detection and with one another at 2 σ level.

Both magnetic field and potential new physics affect the propagation of CMB propagation and generate BB power spectra from EE spectra of CMB. The Faraday rotation due to magnetic field is wavelength dependent while the cosmic polarization rotation due to effective pseudoscalar-photon interaction is wavelength-independent. This property can be used to separate the two effects. With the tensor mode generated by these two effects measured and subtracted, the remaining tensor mode perturbations could be analyzed for signals due to primordial (inflationary) gravitational waves (GWs). In Ni (2009a,b), we have discussed the direct detectability of these primordial GWs using space GW detectors.

**Table 7.** Constraints on cosmic polarization rotation from CMB (cosmic microwave background) polarization observation.

| Reference | Constraint [mrad] | Source data |
|---|---|---|
| Ni (2005a, 2005b) | ±100 | WMAP1 (Bennett *et al* 2003) |
| Feng, Li, Xia, Chen, and Zhang (2006) | −105 ± 70 | WMAP3 (Spergel *et al* 2007) & BOOMERANG (B03) (Montroy *et al* 2006) |
| Liu, Lee, Ng (2006) | ±24 | BOOMERANG (B03) (Montroy *et al* 2006) |
| Kostelecky and Mews (2007) | 209 ± 122 | BOOMERANG (B03) (Montroy *et al* 2006) |
| Cabella, Natoli and Silk (2007) | −43 ± 52 | WMAP3 (Spergel *et al* 2007) |
| Xia, Li, Wang, and Zhang (2008) | −108 ± 67 | WMAP3 (Spergel *et al* 2007) & BOOMERANG (B03) (Montroy *et al* 2006) |
| Komatsu *et al* (2009) | −30 ± 37 | WMAP5 (Komatsu *et al* 2009) |
| Xia, Li, Zhao, and Zhang (2008) | −45 ± 33 | WMAP5 (Komatsu *et al* 2009) & BOOMERANG (B03) (Montroy *et al* 2006) |
| Kostelecky and Mews (2008) | 40 ± 94 | WMAP5 (Komatsu *et al* 2009) |
| Kahniashvili, Durrer, and Maravin (2008) | ± 44 | WMAP5 (Komatsu *et al* 2009) |
| Wu *et al* (2009) | 9.6 ± 14.3 ± 8.7 | QuaD (Pryke *et al* 2009) |

Now we discuss different models for polarization rotation. They all have the same effective Lagrangian and will not be distinguished solely from astrophysical and cosmological polarization observation of polarization rotation. However, some models predict different polarization rotation angles from different observation angles. More precise and comprehensive observations are needed to distinguish models upon detection of cosmic polarization rotation. Feng *et al* (2006) proposed CPT violation and dynamical dark energy. In a more recent paper, Li *et al* (2007) considered baryo/leptogenesis with cosmological CPT violation as a possible cause and gave a 1σ limit on their fermion current-curvature coupling parameter $\delta = -0.011 \pm 0.007$. With the results of Xia, Li, Zhao and Zhang (2008), coupling parameter $\delta$ would be decreased by a factor about 2. Liu, Lee and Ng (2006) gave constraints on the coupling between the quintessence and the pseudoscalar of electromagnetism. Kostelecky and Mews (2007) extended their SME (Standard Model Extension) whose electromagnetic sector is the same as that of χ-g framework with the gravitational constitutive tensor set to constant, to include some higher order terms, and gave constraints on various terms from BOOMERANG. In Kostelecky and Mews (2008), they did their analysis using WMAP5 data. Their most precise constraint is on one of the SME parameter which gives cosmic polarization rotation $\Delta\varphi$.

Geng, Ho and Ng (2007) proposed a new type of effective interactions in terms of the CPT-even dimension-six term equivalent to what we are discussing to generate the cosmic polarization rotation, and used the neutrino number asymmetry to induce a non-zero polarization rotation angle in the data of the CMB polarization. They found that in their model, the rotation effect can be of the order of magnitude of 10-100 mrad or smaller.

In our original pseudoscalar model, the natural coupling strength $\varphi$ is of order 1. However, the isotropy of our observable universe to $10^{-5}$ may leads to a change of $\Delta\varphi$ over cosmological distance scale $10^{-5}$ smaller. Hence, observations to test and measure $\Delta\varphi$ to $10^{-6}$ is very significant. A positive result may indicate that our patch of inflationary universe has a "spontaneous polarization" in the fundamental law of electromagnetic propagation influenced by neighboring patches and by a determination of this fundamental physical law.

The Planck Surveyor was launched in May, 2009. Better sensitivity to $\Delta\varphi$ of $10^{-2}$-$10^{-3}$ (1-10 mrad) is expected. A dedicated future experiment on cosmic microwave background polarization (de Bernardis *et al* 2009, Bock *et al* 2008, KEK 2008) may reach $10^{-5}$-$10^{-6}$ (1-10 μrad) $\Delta\varphi$-sensitivity (Ni 2005a). Astrophysical observations of cosmologically distant objects in various directions will give $\Delta\varphi$ in various directions and will compliment the CMB polarization measurement. Future observations to test and measure $\Delta\varphi$ to $10^{-6}$ and to give $\Delta\varphi$ in various directions are promising.

## 6. Discussion and outlook

During the past decades, we have seen great advances in experiments/observations with orders of magnitude improvement. Up to the present, all the empirical evidences support the Einstein Equivalence Principle and that the intrinsic spin is equivalent to ordinary angular momentum in gravity. With the inflationary-era physics potentially accessible to observations (de Bernardis *et al* 2009, Bock *et al* 2008, KEK 2008, Ni 2009a,b), it is paramount to probe the origin and structure of gravity both experimentally and theoretically. To approach this profound issue, every imaginable way needs to be pursued. To look for polarized photons propagating over astrophysical/cosmological distance is one way, while the measurement of gyrogravitational ratios of elementary particles is another way.

## Acknowledgements

I would like to thank all my collaborators and colleagues for their works that make this review possible. I am grateful to Francis Everitt and Sasha Buchman for helpful discussions on GP-B. I am also grateful to the National Science Council (Grant Nos NSC97-2112-M-007-002 and NSC98-2112-M-007-009) and the National Natural Science Foundation (Grant Nos 10778710 and 10875171) for supporting the writing of this review.**References**